\documentclass[12pt]{article}
\usepackage{amsmath,amssymb,bm}

\newcommand{\B}{\ensuremath{{\cal B}}}
\newcommand{\K}{\ensuremath{{\cal K}}}
\newcommand{\unit}{\ensuremath{\mbox{{\small 1}\hspace*{-0.35em}1}}}
\newcommand{\qed}{~ \rule{0.8ex}{1.6ex}\smallskip}

\newtheorem{theorem}{Theorem}[section]
\newtheorem{lemma}[theorem]{Lemma}
\newtheorem{corollary}[theorem]{Corollary}

\title{Existence of the Spin-Wave Gap in a Deformed Flat-Band Hubbard Model}

\author{Makoto Homma \\{\it Department of Physics, Science and
Technology, Nihon University,}\\ {\it Tokyo, Japan}}

\begin{document}

\maketitle

\begin{abstract}
 We consider a deformed flat-band Hubbard model under a periodic boundary
 condition in arbitrary dimensions. We show that the ground state is
 only all-spin-up or -down state. We obtain upper and lower bounds
 of the one-magnon spin-wave energy with an arbitrary momentum. 
 This dispersion relation is the same as that in the XXZ model 
 in the certain parameter region. Therefore the spin wave has a finite energy 
 gap. These results suggests the our model and the XXZ model.

 \paragraph{Keywords}
 ferromagnetism, flat-band Hubbard model, exact solution, quantum effect, anisotropy effect, spin-wave
\end{abstract}

\section{Introduction}


The origin of the ferromagnetism has been mystery for a long time.
We cannot show the ferromagnetism without quantum many body effect since
interactions for electrons are almost spin independent in the
microscopic point of view. Moreover, only a non-perturbative analysis
can predict the ferromagnetism since perturbative approaches such as the
Fermi liquid theory show only paramagnetism.


Hubbard model is one of the simplest models to describe itinerant
electrons in solids. This model is a lattice electron model which
considered only two effects. One is an 
electron hopping between lattice points and the other is
an on-site repulsive interaction. We identify this interaction with the
Coulomb interaction. 
Recently, Mielke and Tasaki
independently have proposed the models whose ground state has saturated
ferromagnetism by a rigorous constructive approach. These models are called
flat-band Hubbard model \cite{Mielke1,Mielke2,Mielke3,Tasaki92}.
Some remarkable results for ferromagnetic ground states have been
obtained in this class of models. Nishino, Goda and Kusakabe have extended
their result to more general models \cite{NGK}. Tasaki has proved also the
stability of the saturated ferromagnetism against a perturbation which
bends the electron band \cite{Tasaki95,Tasaki03}. Tanaka and Ueda have shown the
stability of the saturated ferromagnetism in a more complicated
two-dimensional model in Mielke's class \cite{TU}. Tasaki has studied
the energy of the spin-wave excitations in the flat-band Hubbard model
\cite{Tasaki96}. He has shown that the dispersion of the one-magnon
excitation is non-singular in the flat-band Hubbard model, contrary to
the Nagaoka ferromagnetism. The flat-band ferromagnetism is believed to
be stable against a small perturbation or change the electron number
density \cite{MT}.

We argue an anisotropy in a ferromagnetic Hubbard model
in this paper. 
A realistic ferromagnet has anisotropies which have no SU(2)
symmetry. The Hubbard model has SU(2) symmetry and
therefore it has no anisotropy. One of the easiest ways
to introduce an anisotropy is adding a spin anisotropy term such as 
the Ising term. However, this way gives us no information for
the origin of the anisotropies as well as the ferromagnetism.
It is believed that the anisotropy is originated from
non-trivial electron dynamics. Moriya has shown
that the spin-orbit coupling gives a spin-dependence of the electron
hopping and the effective spin model has the Ising and the Dzyaloshinski-Moriya
interactions. The hopping anisotropy induces the anisotropy in an effective
spin model. We expect that the spin-orbit coupling induces the
anisotropy even for strong ferromagnetic systems.
%
%
Recently, a deformed flat-band Hubbard model with an exact domain wall
ground state was proposed \cite{HI1,HI2,HI3}. The deformed model has a 
hopping anisotropy which depends on a spin of an electron. 
The ground states of
the model with open boundary condition has same degeneracy as that in
original SU(2) symmetric model. The anisotropy induces the deformation of
SU(2) spin algebra, and then a ground state has a domain wall. It
is also proven that there exists local gapless excitation above the
domain wall ground state \cite{HI3}. These properties of deformed
flat-band Hubbard model are very similar to those in the XXZ model with
critical boundary field. Therefore, we expect that the Ising anisotropy in
quantum spin models comes from hopping anisotropy in 
a ferromagnetic Hubbard model. 


In this paper, we study a deformed flat-band Hubbard model under a
periodic boundary condition. We can construct all exact ground
states. We find that the magnetization of ground states are fully
polarized, {\it i.e.}, the ground states are only two states: all-spin-up
state and -down state. We also obtain upper and lower 
bounds of one-magnon spin-wave
excitation energy. We employ the method proposed by Tasaki. 
He showed that the
spin-wave excitation in the SU(2) invariant 
Tasaki model has the ordinary spin-wave dispersion
relation which has no energy gap above the ground state. We find that
the one-magnon spin-wave energy has the finite gap above the all-up
state in our deformed model. The dispersion relation is
the same as that in XXZ model in the certain
parameter region. These facts indicate that our model is related to XXZ
model which is non-singular spin model.


This paper is organized as follows.
In section \ref{sec:def}, we define a deformed flat-band Hubbard model
and show the main results which consist of three theorems. 
In section \ref{sec:GS}, we prove the first theorem for ground
states. In section \ref{sec:SW_LB}, we show two lemmas and prove 
the second theorem for the lower bound of spin-wave energy from these lemmas. 
In section \ref{sec:calc_ME} and \ref{sec:est_ME},
we prove the lemmas. In section \ref{sec:SW_UB}, we prove the third theorem for the upper
bound of the spin-wave excitation. 

\section{Definitions and Main Results \label{sec:def}}

In this section, we define the $d$-dimensional deformed flat-band
Hubbard model. We also show our main results and discuss their physical
meanings. The proofs of results are given in later sections.

\subsection{Lattice}

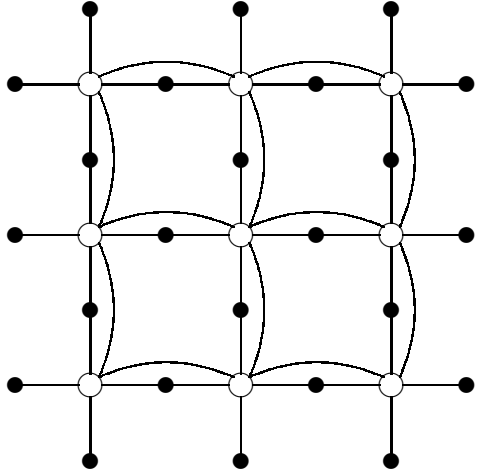
\begin{figure}[htbp]
 \begin{center}
  \setlength{\unitlength}{1mm}
  \begin{picture}(60,60)(-30,-30)
   \put(-20,-30){\circle*{2}}
   \put(-20,-10){\circle*{2}}
   \put(-20,10){\circle*{2}}
   \put(-20,30){\circle*{2}}
   \put(0,-30){\circle*{2}}
   \put(0,-10){\circle*{2}}
   \put(0,10){\circle*{2}}
   \put(0,30){\circle*{2}}
   \put(20,-30){\circle*{2}}
   \put(20,-10){\circle*{2}}
   \put(20,10){\circle*{2}}
   \put(20,30){\circle*{2}}
   \put(-30,-20){\circle*{2}}
   \put(-10,-20){\circle*{2}}
   \put(10,-20){\circle*{2}}
   \put(30,-20){\circle*{2}}
   \put(-30,0){\circle*{2}}
   \put(-10,0){\circle*{2}}
   \put(10,0){\circle*{2}}
   \put(30,0){\circle*{2}}
   \put(-30,20){\circle*{2}}
   \put(-10,20){\circle*{2}}
   \put(10,20){\circle*{2}}
   \put(30,20){\circle*{2}}
   \put(-20,-20){\circle{3}}
   \put(-20,0){\circle{3}}
   \put(-20,20){\circle{3}}
   \put(0,-20){\circle{3}}
   \put(0,0){\circle{3}}
   \put(0,20){\circle{3}}
   \put(20,-20){\circle{3}}
   \put(20,0){\circle{3}}
   \put(20,20){\circle{3}}
   \put(-20,-30){\line(0,1){8.5}}
   \put(0,-30){\line(0,1){8.5}}
   \put(20,-30){\line(0,1){8.5}}
   \put(-20,30){\line(0,-1){8.5}}
   \put(0,30){\line(0,-1){8.5}}
   \put(20,30){\line(0,-1){8.5}}
   \put(-30,-20){\line(1,0){8.5}}
   \put(-30,0){\line(1,0){8.5}}
   \put(-30,20){\line(1,0){8.5}}
   \put(30,-20){\line(-1,0){8.5}}
   \put(30,0){\line(-1,0){8.5}}
   \put(30,20){\line(-1,0){8.5}}
   \put(-18.5,-20){\line(1,0){17}}
   \put(-18.5,0){\line(1,0){17}}
   \put(-18.5,20){\line(1,0){17}}
   \put(1.5,-20){\line(1,0){17}}
   \put(1.5,0){\line(1,0){17}}
   \put(1.5,20){\line(1,0){17}}
   \put(-20,-18.5){\line(0,1){17}}
   \put(0,-18.5){\line(0,1){17}}
   \put(20,-18.5){\line(0,1){17}}
   \put(-20,1.5){\line(0,1){17}}
   \put(0,1.5){\line(0,1){17}}
   \put(20,1.5){\line(0,1){17}}
   \put(20,1.5){\line(0,1){17}}
   \qbezier[100](1.06,21.06)(10,25)(18.93,21.06)
   \qbezier[100](-1.06,21.06)(-10,25)(-18.93,21.06)
   \qbezier[100](1.06,1.06)(10,5)(18.93,1.06)
   \qbezier[100](-1.06,1.06)(-10,5)(-18.93,1.06)
   \qbezier[100](1.06,-18.93)(10,-15)(18.93,-18.93)
   \qbezier[100](-1.06,-18.93)(-10,-15)(-18.93,-18.93)
   \qbezier[100](-18.93,-18.93)(-15,-10)(-18.93,-1.06)
   \qbezier[100](1.06,-18.93)(5,-10)(1.06,-1.06)
   \qbezier[100](21.06,-18.93)(25,-10)(21.06,-1.06)
   \qbezier[100](-18.93,1.06)(-15,10)(-18.93,18.93)
   \qbezier[100](1.06,1.06)(5,10)(1.06,18.93)
   \qbezier[100](21.06,1.06)(25,10)(21.06,18.93)
  \end{picture}
 \end{center}
 \vspace*{-1em}

 \caption{The two dimensional lattice (with $L = 3$). The white circles are
 sites in $\Lambda_o$ and the black dots are sites in
 $\Lambda^\prime$. Electrons at a site can hop to another site if this
 site is connected to the original site with a line or a curve.}
 \label{fig:2-dim_lattice}
\end{figure}

The lattice $\Lambda$ on which defined our deformed Hubbard model is
decomposed into two sublattices
\begin{equation}
 \Lambda = \Lambda_o \cup \Lambda^\prime.
\end{equation}
$\Lambda_o$ is a $d$-dimensional integer lattice  with linear size $L$
defined by
\begin{equation}
 \Lambda_o :=
  \left\{
   x = (x_j)_{j=1}^d \in {\mathbb Z}^{d}
   \biggl| |x_j| \leq \frac{L - 1}{2} \quad j = 1, 2, \cdots, d
  \right\},
\end{equation}
where $(x_j)_{j=1}^d:=(x_1, x_2, \cdots, x_d)$.
$\Lambda^\prime$ can be further decomposed to
$\Lambda_j$ ($j=1, 2, \cdots, d$), {\it i.e.}
\begin{equation}
 \Lambda^\prime = \bigcup_{j = 1}^{d}\Lambda_j.
\end{equation}
$\Lambda_j$ is obtained as a half-integer translation of $\Lambda_o$ to
$j$-th direction,
\begin{equation}
 \Lambda_j := \left\{ x + e^{(j)} | x \in \Lambda_o \right\}
\end{equation}
where $e^{(j)}$ is defined by
\begin{equation}
 e^{(j)} := (\tfrac{1}{2} \delta_{j, l})_{l=1}^d
  = (0, \cdots, 0,
  \begin{array}[t]{@{}l@{}}
   \frac{1}{2}, 0, \cdots, 0 ).\\
   \uparrow\\
   j\mbox{-th}
  \end{array}
\end{equation}
We show the two-dimensional lattice in
Fig. \ref{fig:2-dim_lattice} as an example.

\subsection{Electron Operators and the Fock Space}

The creation and annihilation operators for an electron
$c_{x, \sigma}^\dagger$ and $c_{x, \sigma}$ obey the standard
anticommutation relations
\begin{equation}
 \{ c_{x, \sigma}, c_{y, \tau}^\dagger \}
  = \delta_{x, y} \delta_{\sigma, \tau}, ~~
  \{ c_{x, \sigma}, c_{y, \tau} \}
  = 0 = \{ c_{x, \sigma}^\dagger, c_{y, \tau}^\dagger \},
  \label{eq:anti-comm-rel_for_c}
\end{equation}
where $\{A, B\} = A B + B A$, for $x, y \in \Lambda$ and electron spin coordinates
$\sigma, \tau = \uparrow, \downarrow$.
We define the no-electron state $\Psi_{\rm vac}$ by
\begin{equation}
 c_{x, \sigma} \Psi_{\rm vac} = 0
  \mbox{ for } \forall x \in \Lambda
  \mbox{ and }\sigma = \uparrow, \downarrow.
\end{equation}
We construct a Fock space spanned by a basis
\begin{equation}
 \left\{
  \left( \prod_{x \in A} c_{x, \uparrow}^\dagger \right)
  \left( \prod_{x \in B} c_{x, \downarrow}^\dagger \right)
  \Psi_{\rm vac}
  \biggl| A, B \subset \Lambda
 \right\}.
\end{equation}
We also define a number operator $n_{x, \sigma}$
by $n_{x, \sigma} = c_{x, \sigma}^\dagger c_{x, \sigma}$ whose
eigenvalue represents a number of electrons at site $x$ with spin $\sigma$.
Note anticommutation relations
$\{ c_{x, \sigma}^\dagger, c_{x, \sigma}^\dagger \} = 0$ {\it i.e.}
$c_{x, \sigma}^\dagger c_{x, \sigma}^\dagger = 0.$
This relation implies the Pauli principle. We employ the periodic
boundary condition. This is realized by
$c_{x+2L e^{(j)}, \sigma} = c_{x, \sigma}$ for $j=1,2,\cdots,d$.

\subsection{Deformed Flat-Band Hubbard Model}

The Hubbard model is a model which represents a many-electron system on
an arbitrary lattice. Here, we define a $d$-dimensional deformed
flat-band Hubbard model.
Our model is a generalization of the Tasaki model given in the reference
\cite{Tasaki92}.
A generalized Hubbard Hamiltonian consists of two terms
\begin{equation}
 H := H_{\rm hop} + H_{\rm int}.
  \label{eq:hamiltonian}
\end{equation}
The hopping term $H_{\rm hop}$ is defined by
\begin{align}
 H_{\rm hop} =
 t \sum_{\sigma = \uparrow, \downarrow} \sum_{j=1}^{d}
 \sum_{x \in \Lambda_j}
 d_{x, \sigma}^\dagger d_{x, \sigma}
 \label{eq:H_{hop}}
\end{align}
and the interaction term $H_{\rm int}$ is defined by
\begin{equation}
 H_{\rm int} = U \sum_{x \in \Lambda}
  n_{x, \uparrow} n_{x, \downarrow}
\end{equation}
where $d_{x, \sigma}$ is defined on $x \in \Lambda_j$
$(j=1,2, \cdots, d)$
\begin{equation}
 d_{x, \sigma} :=
  (q^{-p(\sigma)/4})^\ast c_{x - e^{(j)}, \sigma} +
  \lambda c_{x, \sigma} +
  (q^{p(\sigma)/4})^\ast c_{x + e^{(j)}, \sigma},
\end{equation}
with $t, U > 0$. $q$ is an complex parameter and $p(\sigma)$ is defined
by $p(\sigma)=\pm 1$ for $\sigma=\uparrow, \downarrow$.
We write the phase factor of $q$ by $\theta$.
The hopping Hamiltonian $H_{\rm hop}$ can also be written
in the following form
\begin{equation}
 H_{\rm hop} = \sum_{x, y \in \Lambda} t_{x, y}^{(\sigma)}
  c_{x, \sigma}^\dagger c_{y, \sigma}.
\end{equation}
Each term $t_{x, y}^{(\sigma)} c_{x, \sigma}^\dagger c_{y, \sigma}$
in the hopping Hamiltonian represents the hopping of an electron
with spin $\sigma$ from site $x$ to site $y$ with a probability
proportional to $|t_{x, y}^{(\sigma)}|^{2}$.
We expect that the spin-orbit coupling can induce this spin dependent
hopping amplitudes \cite{Moriya}.

Since the interaction Hamiltonian $H_{\rm int}$ represents a on-site
repulsive interaction, we regard this Hamiltonian as a simplification of
the Coulomb interaction between two electrons.

Note that this system conserves the number of electron. The total
electron number operator $\hat{N}_{e}$ is defined by
\begin{equation}
 \hat{N}_{e} := \sum_{x \in \Lambda}
  \sum_{\sigma = \uparrow, \downarrow} n_{x, \sigma}.
\end{equation}
Since the Hamiltonian commutes with this operator, we can set the
electron number to an arbitrary filling. In the
present paper, we consider only the case that the
electron number is equal to $|\Lambda_o|$.
We confine ourselves to the Hilbert space ${\cal H}$
spanned by the following basis
\begin{equation}
 \left\{
  \left( \prod_{x \in A} c_{x, \uparrow}^\dagger \right)
  \left( \prod_{x \in B} c_{x, \downarrow}^\dagger \right)
  \Psi_{\rm vac}
  \biggl| A, B \subset \Lambda \mbox{ with }
  |A| + |B| = |\Lambda_o|
 \right\}.
\end{equation}

Let us discuss the symmetry of the model. An important symmetry is a
U(1) symmetry. We define spin operators at site $x$ by
\begin{equation}
 S_{x}^{(l)} := \sum_{\sigma, \tau = \uparrow, \downarrow}
  c_{x, \sigma}^\dagger
  \frac{{\cal P}_{\sigma, \tau}^{(l)}}{2} c_{x, \tau},
\end{equation}
where  ${\cal P}^{(l)}$ ($l = 1, 2, 3$) denote Pauli matrices
\begin{equation}
 {\cal P}^{(1)} =
  \begin{pmatrix}
   0 & 1 \\ 1 & 0
  \end{pmatrix}
  , \quad
  {\cal P}^{(2)} =
  \begin{pmatrix}
   0 & - i \\ i & 0
  \end{pmatrix}
  , \quad
  {\cal P}^{(3)} =
  \begin{pmatrix}
   1 & 0 \\ 0 & -1
  \end{pmatrix}
  .
\end{equation}
The Hamiltonian commutes with the third component of total spin operator
\begin{equation}
 [H, S_{\rm tot}^{(3)}] =
  H S_{\rm tot}^{(3)} - S_{\rm tot}^{(3)} H = 0,
\label{good}
\end{equation}
with
\begin{equation}
 S_{\rm tot}^{(l)} = \sum_{x \in \Lambda} S_{x}^{(l)}.
\end{equation}
Note that this symmetry is enhanced to an SU(2) symmetry in the case of
$q = 1$ {\it i.e.} Hamiltonian commutes with any component of total spin
operator.
In this case, this model becomes the original flat-band Hubbard
model given by Tasaki \cite{Tasaki92,Tasaki95}.
Another important symmetry is ${\mathbb Z}_2$ symmetry which is
generated by a product of a parity and spin rotation defined by
\begin{equation}
 \Pi = \Pi^{-1} = P \exp \left( i \pi S_{\rm tot}^{(1)} \right),
\end{equation}
where $P$ is a parity operator defined by
$P c_{x, \sigma} P = c_{-x, \sigma}$ and
$P c_{x, \sigma}^\dagger P = c_{-x, \sigma}^\dagger$.
$\Pi$ transforms $c_{x, \sigma}$ and $c_{x, \sigma}^\dagger$ to
$c_{-x, \overline{\sigma}}$ and $c_{-x, \overline{\sigma}}^\dagger$,
where $\overline{\sigma} = \downarrow$ if $\sigma = \uparrow$ or
$\overline{\sigma} = \uparrow$ if $\sigma = \downarrow$.
Note the following transformation of the total magnetization
$\Pi S_{\rm tot}^{(3)} \Pi = - S_{\rm tot}^{(3)}$. An energy eigenstate
with the total magnetization $M$ is transformed by $\Pi$ into another
eigenstate with the total magnetization $-M$, which belongs to the same
energy eigenvalue.

\subsection{Ground States}

First, we introduce two states $\Psi_\uparrow$ and $\Psi_\downarrow$
defined by
\begin{equation}
 \Psi_\uparrow =
  \left(\prod_{x \in \Lambda_o} a_{x, \uparrow}^\dagger \right)
  \Psi_{\rm vac}
  \mbox{ and }
 \Psi_\downarrow =
  \left(\prod_{x \in \Lambda_o} a_{x, \downarrow}^\dagger \right)
  \Psi_{\rm vac},
\end{equation}
where $a_{x, \sigma}^\dagger$ with $x \in \Lambda_o$ is
\begin{equation}
 a_{x, \sigma} = - q^{p(\sigma)/4} \sum_{j = 1}^{d}
   c_{x - e^{(j)}, \sigma}^\dagger
   + \lambda c_{x, \sigma}^\dagger
   - q^{-p(\sigma)/4} \sum_{j = 1}^{d}
   c_{x + e^{(j)}, \sigma}^\dagger.
\end{equation}
We can easily verify that the above two states are ground states of the
model since the Hamiltonian is positive semi-definite and
$[d_{y, \sigma}, a_{x, \tau}^\dagger]=0$ for $y \in \Lambda^\prime$ and
$x \in \Lambda_o$. The following theorem states that the ground states 
of the model are only above two states.

{\theorem(Ground State of the Model)
In the deformed flat-band Hubbard model defined by
(\ref{eq:hamiltonian}) under a periodic boundary condition with a fixed electron
number $|\Lambda_o|$, the space of ground states are spanned by
two fully polarized states $\Psi_\uparrow$ and $\Psi_\downarrow$.
\label{th:GS}}\smallskip

The proof is given in the next section.

This theorem shows that the deformation of electron hopping destroys
the large degeneracy of ground state in the original SU(2) symmetric
Tasaki model and we obtain two ground states: all-up state and all-down
state. The ground states breaks ${\mathbb Z}_2$ symmetry.
This is similar to that Ising anisotropy effect of the Ising-like XXZ
model.

\subsection{Spin-Wave Excitations}

Before we discuss the spin-wave excitation, we remark properties of the
spin-wave state in quantum spin models. The one-magnon spin-wave state $\Phi_{\rm SW}(k)$
with a wave-number $k \in \K$ satisfies
\begin{equation}
 T_{x} \Phi_{\rm SW}(k) = e^{- i k \cdot x} \Phi_{\rm SW}(k)
  \label{eq:SW_1}
\end{equation}
and
\begin{equation}
 S_{\rm tot}^{(3)} \Phi_{\rm SW}
  = (S_{\rm max} - 1 ) \Phi_{\rm SW},
  \label{eq:SW_2}
\end{equation}
where $x \in \Lambda_o$ and the translation operator $T_x$ is
defined by
\begin{equation}
 T_x c_{y, \sigma} T_x^{-1} = c_{x + y, \sigma}
  \quad \mbox{and} \quad
 T_x c_{y, \sigma}^\dagger T_x^{-1} = c_{x + y, \sigma}^\dagger.
\end{equation}
$\K$ is the space of wave number vectors
\begin{equation}
 \K :=
  \left\{
  \frac{2 \pi n}{L} ~\biggl|~
  n \in {\mathbb Z}^d \cap \left[ -\frac{L-1}{2}, \frac{L-1}{2} \right]^d
  \right\}.
  \label{eq:def_of_K}
\end{equation}
Then, the one-magnon spin wave state should be in the following Hilbert
space ${\cal H}_k$
\begin{equation}
 {\cal H}_k :=
  \left\{
   \Psi \in {\cal H} ~\Bigl|~
   T_x \Psi = e^{-i k \cdot x} \Psi \mbox{ and }
   S_{\rm tot}^{(3)} \Psi = \tfrac{1}{2} (|\Lambda_o| - 1) \Psi
  \right\}.
\end{equation}
We define one-magnon spin-wave state as the lowest energy state
in ${\cal H}_k$. We denote the spin-wave energy with wave-number $k$ by
$E_{\rm SW}(k)$. We can show the following two theorems. Note that
$\theta \in {\mathbb R}$ which appears in followings is the phase factor of $q$,
{\it i.e.} $q=|q|e^{i\theta}$.

{\theorem(Lower Bound of the Spin-Wave Excitation)
There exist positive constants $t_1$, $\lambda_1$, $U_1$ and $A_1$
independent of the system size such that
\begin{equation}
 E_{\rm SW}(k) \geq
  \frac{2 U}{\lambda^{4}}
  \left[
   \frac{d(|q| + |q|^{-1})}{2}
   - \sum_{j = 1}^{d} \cos \left( 2 k \cdot e^{(j)} + \theta \right)
   - \frac{A_1}{\lambda}
  \right]
\end{equation}
for $t \geq t_1$, $\lambda \geq \lambda_1$ and $U \geq U_1$.
\label{th:SW_LB}}

{\theorem(Upper Bound of the Spin-Wave Excitation)
There exists positive constant $A_2$ such that
\begin{equation}
  E_{\rm SW}(k) \leq
  \frac{2 U}{\lambda^{4}}
  \left[
   \frac{d(|q| + |q|^{-1})}{2}
   - \sum_{j = 1}^{d} \cos \left( 2 k \cdot e^{(j)} + \theta \right)
   + \frac{A_2}{\lambda^2}
  \right]
\end{equation}
for $t, \lambda, U>0$.
\label{th:SW_UB}}

\paragraph{Remark}

Theorems \ref{th:SW_LB} and \ref{th:SW_UB} show that the dispersion
relation of the spin-wave is asymptotically
\begin{equation}
   E_{\rm SW}(k) \sim
  \frac{2 U}{\lambda^{4}}
  \left[
   \frac{d(|q| + |q|^{-1})}{2}
   - \sum_{j = 1}^{d} \cos \left( 2 k \cdot e^{(j)} + \theta \right)
  \right]
  \label{eq:SW_asymp}
\end{equation}
for large $\lambda$. This dispersion relation is the same as that in the
XXZ model with the Dzyaloshinski-Moriya interaction given by the
Hamiltonian
\begin{equation}
 -J \sum_{x \in \Lambda_o} \sum_{j=1}^d
  \left[
   \cos \theta
   \left( S_x^{(1)} S_{x+2e^{(j)}}^{(1)}
   +
   S_x^{(2)} S_{x+2e^{(j)}}^{(2)} \right)
   + \Delta S_x^{(3)} S_{x+2e^{(j)}}^{(3)}
   -\sin \theta {\bm D}\cdot ({\bm S}_x \times {\bm S}_{x+2e^{(j)}})
  \right]
  \label{eq:H_eff}
\end{equation}
with an exchange parameter $J=2U/\lambda^4$, an Ising anisotropy
parameter $\Delta=(|q|+|q|^{-1})/2$ and a Dzyaloshinski-Moriya
vector ${\bm D} = (0,0,1)$. Furthermore we can obtain the
representation (\ref{eq:H_eff}) as an effective Hamiltonian by a
perturbation theory from our model with a perturbation parameter
$1/\lambda$.

{\corollary(Existence of the Spin-Wave Gap)
There exists positive constant $\lambda_3$ such that
\begin{equation}
 \min_{k \in {\cal K}} E_{\rm SW}(k) \geq
  \frac{2 U}{\lambda^{4}}
  \left[
   \frac{d(|q| + |q|^{-1}-2)}{2}
   - \frac{A_1}{\lambda}
  \right]
  > 0
\end{equation}
\label{cor:SW_Gap}}

This means that there is a finite energy gap between 
the fully polarized ground state and a spin flipped state. We expect that the
energy spectra have finite gap above the ground states. On the contrary,
there exists a gapless excitation in the model under an open boundary
condition. This drastic difference is also similar to the XXZ model.

\section{Ground States\label{sec:GS}}

In this section, we obtain ground states with the fixed
electron number $N_{e} = |\Lambda_o|$,
and prove the Theorem \ref{th:GS} on the basis of Tasaki's
construction method \cite{Tasaki92}.

\subsection{Localized Electron Operators}

First, We introduce localized electron operators, which are convenient to
construct a ground state and to prove the bounds of spin-wave
excitation energy. This representation was introduced by Tasaki\cite{Tasaki96} to
construct the bases of single electron state. We show the construction
of the operators in Appendix\ref{A:localized_bases}.

First we introduce a localized electron operator $a_{x, \sigma}^\dagger$
defined by
\begin{equation}
 a_{x, \sigma}^\dagger = \sum_{ y \in \Lambda}
  \psi_{y, \sigma}^{(x)} c_{y, \sigma}^\dagger,
  \label{eq:def_of_a}
\end{equation}
where $\psi_{y, \sigma}^{(x)}$ is represented by
\begin{equation}
 \psi_{y, \sigma}^{(x)} =
  \begin{cases}
   \displaystyle
   -\frac{q^{p(\sigma)/4}}{\lambda}
   \sum_{j=1}^{d} \delta_{x - e^{(j)}, y}
   + \delta_{x, y}
   -\frac{q^{-p(\sigma)/4}}{\lambda}
   \sum_{j=1}^{d} \delta_{x + e^{(j)}, y}
   & \mbox{for }x \in \Lambda_o \\
   \displaystyle
   \frac{(q^{-p(\sigma)/4})^\ast}{\lambda}
   \delta_{x - e^{(j)}, y}
   + \delta_{x, y}
   + \frac{(q^{p(\sigma)/4})^\ast}{\lambda}
   \delta_{x + e^{(j)}, y}
   & \mbox{for }x \in \Lambda_j
  \end{cases}
  .
  \label{eq:local_basis}
\end{equation}
$a_{x, \sigma}^\dagger \Psi_{\rm vac}$ with $x \in \Lambda_o$ is a
ground state of single electron state. The single-electron ground state
has
$|\Lambda_o|$-fold degeneracy. This macroscopic degeneracy of
single-electron ground state is one of the origin of flat-band
ferromagnetism.
$\{ a_{x, \sigma}^\dagger \Psi_{\rm vac} \}_{x \in \Lambda_o}$ is 
a basis of zero-energy single-electron states, and 
$\{ a_{y, \sigma}^\dagger \Psi_{\rm vac} \}_{y \in \Lambda^\prime}$
is a basis of excited single-electron states. Furthermore, the ground state energy is 0 and the
lowest excitation energy eigenvalue is $t \lambda^2$ (See
Appendix\ref{A:localized_bases}).

We can also construct another localized electron operator
$b_{x, \sigma}$ which satisfies
\begin{equation}
 \{ b_{x, \sigma}, a_{y, \tau}^\dagger \} = \delta_{x, y}
  \delta_{\sigma, \tau}
  \mbox{ and }
 \{ b_{x, \sigma}, b_{y, \tau} \}
 = 0 =
 \{ a_{x, \sigma}^\dagger, a_{y, \tau}^\dagger \}.
 \label{eq:a-com_a-b}
\end{equation}
We represent $b_{x, \sigma}$ in terms of the original electron operator
by
\begin{equation}
 b_{x, \sigma} = \sum_{y \in \Lambda}
  (\tilde{\psi}_{y, \sigma}^{(x)})^\ast c_{y, \sigma}.
  \label{eq:def_of_b}
\end{equation}
(\ref{eq:def_of_a}), (\ref{eq:def_of_b}) and (\ref{eq:a-com_a-b}) means
\begin{equation}
 \sum_{w \in \Lambda}
  (\tilde{\psi}_{w, \sigma}^{(x)})^\ast \psi_{w, \sigma}^{(y)}
  = \delta_{x, y}
  \mbox{ and }
  \sum_{w \in \Lambda}
  (\tilde{\psi}_{x, \sigma}^{(w)})^\ast \psi_{y, \sigma}^{(w)}
  = \delta_{x, y}.
  \label{eq:dual_basis}
\end{equation}
Note, $\tilde{\psi}_{y, \sigma}^{(x)}$ cannot be represented closed form
but decay exponentially as $\|x-y\|_1$ becomes large, where
$\|x-y\|_1:=\sum_{j=1}^{d}|x_j-y_j|$. We can obtain the following
bounds.

{\lemma
There exists positive constant $B_1$ such that
\begin{align}
 & \sum_{w \in \Lambda} |\tilde{\psi}_{w, \sigma}^{(x)}-\psi_{w, \sigma}^{(x)}|
 \leq \frac{B_1}{\lambda^2},
 \label{eq:bound_of_sum_psi-1}
 \\
 & \sum_{w \in \Lambda} |\tilde{\psi}_{x, \sigma}^{(w)}-\psi_{x, \sigma}^{(w)}|
 \leq \frac{B_1}{\lambda^2}
 \label{eq:bound_of_sum_psi-2}
\end{align}
for all $x, y \in \Lambda$.
\label{lemma:bound_sum_psi}}

{\lemma
There exist positive constants $B_2$, $B_3$ and $B_4$ 
\begin{align}
 & |\tilde{\psi}_{y, \sigma}^{(x)}|
 \leq B_2
 \left[
 \frac{2d}{\lambda^2 +d (|q|^{1/2}+|q|^{-1/2})}
 \right]^{\|y-x\|_1}
 \mbox{ for } \forall x, y \in \Lambda,
 \label{eq:bound_of_psi} \\
 & |\tilde{\psi}_o^{(o)} - \psi_o^{(o)}|
 \leq
 \frac{B_3}{\lambda^2},
 \label{eq:bound_diff_psi00} \\
 \intertext{and}
 & |\tilde{\psi}_{\pm e^{(j)}}^{(o)} - \psi_{\pm e^{(j)}}^{(o)}|
 \leq
 \frac{B_4}{\lambda^3}.
 \label{eq:bound_diff_psi01}
\end{align}
\label{lemma:bound_psi}}

The proofs of these lemmas are given in \ref{A:estimates_of_psi}.

The original electron operators can be written in terms of
$a_{x, \sigma}^\dagger$ and $b_{x, \sigma}$,
\begin{equation}
 c_{x, \sigma}^\dagger
  = \sum_{y \in \Lambda}
  (\tilde{\psi}_{x, \sigma}^{(y)})^\ast a_{y, \sigma}^\dagger
  \quad \mbox{and} \quad
  c_{x, \sigma}
  = \sum_{y \in \Lambda} \psi_{x, \sigma}^{(y)} b_{y, \sigma}.
  \label{eq:c_i.t.o._a_or_b}
\end{equation}
The Hilbert space with $|\Lambda_o|$ electrons is also spanned by the
basis
\begin{equation}
 \left\{
  \left( \prod_{x \in A} a_{x, \uparrow}^\dagger \right)
  \left( \prod_{x \in B} a_{x, \downarrow}^\dagger \right)
  \Psi_{\rm vac}
  \biggl| A, B \subset \Lambda \mbox{ with }
  |A| + |B| = |\Lambda_o|
 \right\},
\end{equation}
because $c_{x, \sigma}^\dagger$ can be written in terms of
$a_{x, \sigma}^\dagger$.

\subsection{Proof of Theorem \ref{th:GS}}

Here, we construct ground states of the model.
Let $\Psi$ be a ground state with $|\Lambda_0|$ electrons. First, we
expand a ground state $\Psi$ into the following series
\begin{equation}
 \Psi = \sum_{A, B}
  \psi(A, B)
  \left(
   \prod_{x \in A} a_{x, \uparrow}^\dagger
  \right)
  \left(
   \prod_{y \in B} a_{y, \downarrow}^\dagger
  \right)
  \Psi_{\rm vac} \label{eq:exp-of-GS-1},
\end{equation}
where the summation is taken over all $A, B \subset \Lambda$ with
$|A|+|B|=|\Lambda_o|$. Note that $a_{x, \sigma}^\dagger$ for
$x \in \Lambda_o$ creates an electron which has the lowest energy of
$H_{\rm hop}$. For the lowest energy state of
$H_{\rm hop}$, $\psi(A, B)$ should vanish, if $A$ or $B$
contains a site in $\Lambda^\prime$.
Next, we consider the interaction Hamiltonian. If we find a state $\Psi$
such that $c_{x, \uparrow} c_{x, \downarrow} \Psi=0$ for
$\forall x \in \Lambda$, then the state is a ground state of
$H_{\rm int}$. For $x \in \Lambda_o$, we find that $\psi(A, B)$ survives
only for $A \cap B = \emptyset$, if
$A, B \subset \Lambda_o$. These facts allow us to represent $\Psi$ the
following form:
\begin{equation}
 \Psi =
  \sum_\sigma \phi(\sigma)
  \left(
   \prod_{x \in \Lambda_o} a_{x, \sigma_{x}}^\dagger
  \right) \Psi_{\rm vac}
  \label{eq:rep_of_GS_M}
\end{equation}
where the summation is taken over all possible spin configurations
$\sigma=(\sigma_{x})_{x \in \Lambda_o}$.
To satisfy the condition $c_{y, \uparrow} c_{y, \downarrow} \Psi=0$ for
$y \in \Lambda_j$ ($j=1,2,\cdots, d$), the coefficient holds
\begin{equation}
 \phi(\sigma) = q^{\left[ p(\sigma_{y - e^{(j)}})
  - p(\sigma_{y + e^{(j)}}) \right]/2}
  \phi(\sigma_{y - e^{(j)}, y + e^{(j)}}),
  \label{eq:cond_for_phi}
\end{equation}
where $\sigma_{x, y}$ is spin configuration obtained by the exchange
$\sigma_{x}$ and $\sigma_{y}$ in the original configuration $\sigma$.
The periodic boundary
condition allows no configuration which satisfies the condition
(\ref{eq:cond_for_phi}) except in the two cases: $\sigma_x = \uparrow$
for all $x \in \Lambda_o$ or $\sigma_x = \downarrow$ for all
$x \in \Lambda_o$. Thus, we conclude that all ground states in the
periodic boundary condition are only two fully polarized states
\begin{equation}
 \Psi_\uparrow :=
  \left( \prod_{x \in \Lambda_o} a_{x, \uparrow}^\dagger \right)
  \Psi_{\rm vac}
\end{equation}
and
\begin{equation}
 \Psi_\downarrow :=
  \left( \prod_{x \in \Lambda_o} a_{x, \downarrow}^\dagger \right)
  \Psi_{\rm vac}.
\end{equation}
This fact proves Theorem \ref{th:GS}. \qed

\section{Lower bound of the Spin-Wave Excitation \label{sec:SW_LB}}

In this section, we show that the one-magnon spin-wave excitation has the
same dispersion relation as that in the XXZ model. Our proof is based on
Tasaki's argument for the SU(2) invariant model \cite{Tasaki96}.	He has
proved that the one-magnon spin-wave excitation in the Tasaki model has
the same dispersion relation as that in the ferromagnetic Heisenberg
model.	These spin-wave excitations in both models have no energy gap,
since they are the Goldstone mode above the ground state which
spontaneously breaks the SU(2) spin rotation symmetry. On the contrary,
we will find an energy gap in our anisotropic model.

\subsection{Another Hopping Hamiltonian}

To estimate a lower bound, it is convenient to introduce a
new hopping Hamiltonian $\tilde{H}_{\rm hop}$ which satisfies
\begin{equation}
 \tilde{H}_{\rm hop} a_{x, \sigma}^\dagger \Psi_{\rm vac} =0
  \mbox{ for } x \in \Lambda_o
\end{equation}
and
\begin{equation}
 \tilde{H}_{\rm hop} a_{x, \sigma}^\dagger \Psi_{\rm vac} =
  t \lambda^2 a_{x, \sigma}^\dagger \Psi_{\rm vac}
  \mbox{ for } x \in \Lambda^\prime.
\end{equation}
Note that $a_{x, \sigma}^\dagger \Psi_{\rm vac}$ with $x \in \Lambda_o$
is the basis of single electron  ground states 
and $a_{y, \sigma}^\dagger \Psi_{\rm vac}$ with $y \in \Lambda^\prime$ is
a basis of single electron 
excited states. Moreover, the ground state energy
is 0 and the lowest excitation energy eigenvalue is $t \lambda^2$, {\it i.e.}
$H_{\rm hop} a_{x, \sigma}^\dagger \Psi_{\rm vac}=0$ for all
$x \in \Lambda_o$ and
$(\Psi, H_{\rm hop} \Psi)/\|\Psi\|^2 \geq t \lambda^2$ with
$\Psi = \sum_{y \in \Lambda^\prime} C_y a_{y, \sigma}^\dagger \Psi_{\rm vac}$
for any set of complex numbers $\{ C_y \}_{y \in \Lambda^\prime}$.
Then, we obtain that $\tilde{H}_{\rm hop} \leq H_{\rm hop}$.
$\tilde{H}_{\rm hop}$ is represented by
\begin{equation}
 \tilde{H}_{\rm hop} = t \lambda^2
  \sum_{x \in \Lambda^\prime} \sum_{\sigma=\uparrow,\downarrow}
  a_{x, \sigma}^\dagger b_{x, \sigma}.
\end{equation}
Note that $\tilde{H}_{\rm hop}$ is not hermitian and therefore 
its eigenvectors are not orthogonal systems.
Nevertheless, all eigenvalues of $\tilde{H}_{\rm hop}$ are real. 
We define a new Hamiltonian
$\tilde{H} := \tilde{H}_{\rm hop}+H_{\rm int}$.

We use a representation of interaction Hamiltonian in terms of the
localized electron operators to evaluate the bounds
\begin{align}
 H_{\rm int} = \sum_{x_1, x_2, x_3, x_4 \in \Lambda}
 \widetilde{U}_{x_1, x_2; x_3, x_4}
 a_{x_1, \uparrow}^\dagger a_{x_2, \downarrow}^\dagger
 b_{x_3, \downarrow} b_{x_4, \uparrow},
\end{align}
where $\tilde{U}_{x_1, x_2; x_3, x_4}$ is defined by
\begin{equation}
 \tilde{U}_{x_1, x_2; x_3, x_4}
  :=
  U \sum_{w \in \Lambda}
  (\tilde{\psi}_{w, \uparrow}^{(x_1)}
  \tilde{\psi}_{w, \downarrow}^{(x_2)})^\ast
  \psi_{w, \downarrow}^{(x_3)} \psi_{w, \uparrow}^{(x_4)}
  .
\end{equation}

\subsection{Basis for the Spin-Wave State}

To define convenient basis of ${\cal H}_k$, we define a state
$\Psi_{\mu, A}(k)$ for $\mu=0,1, \cdots, d$ and for a set
$A \subset \Lambda$ with $|A| = |\Lambda_o|-1$ by
\begin{equation}
 \Psi_{\mu, A}(k) :=
  \sum_{w \in \Lambda_o} e^{i k \cdot w} T_w
  a_{e^{(\mu)}, \downarrow}^\dagger
  \left(
   \prod_{v \in A} a_{v, \uparrow}^\dagger
  \right)
  \Psi_{\rm vac},
\end{equation}
where $e^{(\mu)}=o=(0,0,\cdots, 0)$ for $\mu=0$ and $e^{(\mu)}=e^{(j)}$
for $\mu=j$ $(j=1, 2, \cdots, d)$.
This state satisfies both properties (\ref{eq:SW_1}) and
(\ref{eq:SW_2}). We define another state $\Omega(k)$ by
\begin{equation}
 \Omega(k) = \frac{1}{\alpha}
  \sum_{w \in \Lambda_o} e^{i k \cdot w} T_w
  a_{o, \downarrow}^\dagger b_{o, \uparrow} \Psi_\uparrow(k)
  \propto \Psi_{0, \Lambda_o \backslash \{ o \}},
\end{equation}
which is an approximation of the spin wave state. We will choose a
positive constant $\alpha$ in the proof. We define the following basis
of ${\cal H}_k$ by
\begin{align}
 \B_{k} :=
 \{ \Omega(k) \} \cup
 \bigl\{
 \Psi_{\mu, A}(k)
 ~\bigl|~
 \mu = 0, 1, \cdots, d, \quad A \subset \Lambda
 & \nonumber \\
 \mbox{ with }
 |A| = |\Lambda_o| -1
 \mbox{ and }
 (\mu, A) \neq (0, \Lambda_o \backslash \{ o \})
 &
 \bigr\}.
\end{align}

\subsubsection{Basic Lemma for Lower Bound}

We define matrix elements $h[\Psi, \Phi]$ between $\Psi, \Phi \in \B_k$
by the unique expansion
\begin{equation}
 \tilde{H} \Psi = \sum_{\Phi \in \B_{k}}
  h[\Phi, \Psi] \Phi.
  \label{eq:def_ME}
\end{equation}
And we define $D[\Phi]$ by
\begin{equation}
 D[\Phi] := \Re[h[\Phi, \Phi]]
  - \sum_{\Psi \in \B_k \backslash \{ \Phi \}}
  |h[\Phi, \Psi]|.
\end{equation}
We can prove the following lemma. The proof is given in
Appendix\ref{A:proof_SW-1}.

{\lemma \label{lemma:SW-1}
Let $\tilde{E}_0 (k)$ be the lowest energy eigenvalue of $\tilde{H}$ in
the Hilbert space ${\cal H}_k$. Then, we have
\begin{equation}
 \tilde{E}_0 (k) \geq \min_{\Phi \in \B_k} D[\Phi].
\end{equation}
}

\subsection{Proof of Theorem \ref{th:SW_LB}}

First, we show the results of estimates. We obtain these bounds from the
direct evaluation (See Section \ref{sec:calc_ME} and \ref{sec:est_ME})

{\lemma
There exist positive constants $F_1$, $F_2$, $F_3$, $F_4$, $F_5$ and $F_6$
such that
\begin{align}
 D[\Omega(k)] \geq &
  \frac{2 U}{\lambda^4}
  \left[
   \frac{d (|q|+|q|^{-1})}{2}- \sum_{j=1}^d\cos(k \cdot 2e^{(j)} + \theta)
   - \frac{F_1}{\lambda^2}
  \right]
  - 
 U \alpha \frac{F_2}{\lambda}
 \label{eq:D[Omega]}
 \\
 D[\Psi_{0, x}(k)] \geq &
 U
 \left(
 1 - \frac{F_2}{\lambda} - \frac{F_3}{\lambda^2}
 \right)
 -\frac{U}{\alpha} \frac{F_4}{\lambda^4}
 \label{eq:D[Psi_0x]}
 \\
 D[\Psi_{\mu, A}(k)] \geq &
 t \lambda^2 - U \frac{F_5}{\lambda}
 -\frac{U}{\alpha} \frac{F_6}{\lambda^2},
 \label{eq:D[Psi_gen]}
\end{align}
for $\mu\neq 0$ and $A \cap \Lambda^\prime \neq \emptyset$.
\label{lemma:D[Psi]}}\smallskip

We give the proof of this lemma in subsection
\ref{sec:estimates_of_D[Psi]}.

\paragraph{Proof of Theorem \ref{th:SW_LB} from Lemma \ref{lemma:SW-1} and \ref{lemma:D[Psi]}}

From these inequalities and good choice of $\alpha$,
$\min_{\Psi \in \B_k} D[\Psi]$ is given by $D[\Omega(k)]$.
If we choose $\alpha= U/\lambda^4$, then we find
\begin{align}
 D[\Psi_{0, x}] \geq &
 U
 \left(
 1 - \frac{F_2}{\lambda} - \frac{F_3}{\lambda^2}
 \right)
 -F_4
 = 
 U
 \left( 1 - \frac{F_2+F_3/\lambda}{\lambda} \right)
 -F_4
\end{align}
and
\begin{equation}
 D[\Psi_{\mu, A}] \geq
  t \lambda^2
  \left(
   1- \frac{F_6}{t}
  \right)
  - U \frac{F_5}{\lambda}.
\end{equation}
If $\lambda \geq \lambda_2 = 4 (F_2+F_3/\lambda_2)$ and $U \geq U_1 = 4F_4$, then
\begin{equation}
 D[\Psi_{0, x}] \geq \frac{U}{2}.
\end{equation}
If $t \geq t_1 = 4F_6$ and
$\lambda \geq \lambda_3 = [U F_5/(4t)]^{1/3}$, then
\begin{equation}
 D[\Psi_{\mu, A}] \geq \frac{t \lambda^2}{2}.
\end{equation}

If
$\lambda \geq \lambda_4 = \max \left\{ [d(|q|+|q|^{-1}+2)]^{1/4},[d U(|q|+|q|^{-1}+2)/t]^{1/6}\right\}$,
then
\begin{equation}
 D[\Omega] \leq \frac{d U}{\lambda^4}(|q|+|q|^{-1}+2)
  \leq \frac{U}{2}
\end{equation}
and
\begin{equation}
 D[\Omega] \leq \frac{d U}{\lambda^4}(|q|+|q|^{-1}+2)
  \leq \frac{t \lambda^2}{2}.
\end{equation}
We set $\lambda_1 = \max \{ \lambda_2, \lambda_3, \lambda_4 \}$.
Thus we obtain
\begin{align}
 \min_{\Phi \in \B_{k}} D[\Phi] = D[\Omega(k)]
 \geq
  \frac{2 U}{\lambda^{4}}
  \left[
   \frac{d(|q| + |q|^{-1})}{2}
   - \sum_{j = 1}^{d} \cos \left( 2 k \cdot e^{(j)} + \theta \right)
   - \frac{A_1}{\lambda}
  \right]
\end{align}
for $\lambda \geq \lambda_1$, $U \geq U_1$ and $t \geq t_1$
where $A_1=F_2+F_1/\lambda_1$.
Lemma \ref{lemma:SW-1} and this show that
\begin{equation}
 \tilde{E}_0(k) \geq
  \frac{2 U}{\lambda^{4}}
  \left[
   \frac{d(|q| + |q|^{-1})}{2}
   - \sum_{j = 1}^{d} \cos \left( 2 k \cdot e^{(j)} + \theta \right)
   - \frac{A_1}{\lambda}
  \right].
\end{equation}
Since $H \geq \tilde{H}$, then $E_{\rm SW}(k) \geq \tilde{E}_0(k)$.
Thus we obtain
\begin{equation}
 E_{\rm SW}(k) \geq 
  \frac{2 U}{\lambda^{4}}
  \left[
   \frac{d(|q| + |q|^{-1})}{2}
   - \sum_{j = 1}^{d} \cos \left( 2 k \cdot e^{(j)} + \theta \right)
   - \frac{A_1}{\lambda}
  \right].
\end{equation}
This concludes Theorem \ref{th:SW_LB}. \qed

\section{Calculation of Matrix Elements \label{sec:calc_ME}}

In this section, we calculate the matrix elements defined by
(\ref{eq:def_ME}) to estimate $D[\Phi]$ for $\Phi \in \B_k$. Before 
showing the details of calculation, we list the summary formulae of matrix
elements. We often abbreviate the $k$-dependence of the states for
simplicity.

\subsection{Summary of Matrix Elements}

Here, we show only summary of the calculations. We define
$h_{\rm hop}[\Phi, \Psi]$ and $h_{\rm int}[\Phi, \Psi]$ by
\begin{equation}
 \tilde{H}_{\rm hop} \Psi
  = \sum_{\Phi \in \B_k} h_{\rm hop}[\Phi, \Psi] \Phi
\end{equation}
and
\begin{equation}
 H_{\rm int} \Psi
  = \sum_{\Phi \in \B_k} h_{\rm int}[\Phi, \Psi] \Phi.
\end{equation}
It is convenient to define $\Psi_{\mu, x}$ and $\Psi_{\mu, y, v, w}$
with $x, v, w \in \Lambda_o$, $v \neq w$ and $y \in \Lambda^\prime$ by
\begin{equation}
 \Psi_{\mu, x}(k) :=
  \sum_{w_1 \in \Lambda_o}
  e^{i k \cdot w_1}
  a_{w_1+e^{(\mu)}, \downarrow}^\dagger b_{w_1+x, \uparrow}
  \Phi_\uparrow
  \propto
  \Psi_{\mu, \Lambda_o\backslash\{x\}}
  .
\end{equation}
and
\begin{equation}
 \Psi_{\mu, y, v, w}(k) :=
  \sum_{w_1 \in \Lambda_o}
  e^{i k \cdot w_1}
  a_{w_1+e^{(\mu)}, \downarrow}^\dagger
  b_{v+w_1, \uparrow} b_{w+w_1, \uparrow}
  a_{y+w_1, \uparrow}^\dagger
  \Phi_\uparrow
  \propto
  \Psi_{\mu, (\Lambda_o\backslash\{v, w\})\cup\{y\}}.
\end{equation}
Only $\Psi_{\mu, x}$ and $\Psi_{\mu, y, v, w}$ contribute to
matrix elements related to $\Omega$: $h_{\rm int}[\Omega, \Phi]$ and
$h_{\rm int}[\Phi, \Omega]$.

We summarize the calculations of the matrix elements.
\begin{equation}
 h_{\rm hop}[\Psi_{\rho, B}, \Psi_{\mu, A}] = t \lambda^2
 (|A\cap \Lambda^\prime| + \chi[\mu \neq 0])\delta_{\mu, \rho} \chi[A=B],
 \label{eq:h_hop}
\end{equation}
\begin{align}
 h_{\rm int}[\Psi_{\rho, y}, \Psi_{\mu, x}]
  =
 e^{-i k \cdot x}
  \sum_{v \in \Lambda_o}
  \left(
   e^{i k \cdot y}
   \tilde{U}_{v, e^{(\rho)}; y-x+e^{(\mu)}, v}
   - e^{i k \cdot v}
   \tilde{U}_{v, e^{(\rho)}; v-x+e^{(\mu)}, y}
  \right),
  \label{eq:h_int[1]}
\end{align}
\begin{equation}
 h_{\rm int}[\Psi_{\rho, y, v, w}, \Psi_{\mu, x}]
  = 
  e^{i k \cdot (v-x)} \tilde{U}_{y, e^{(\rho)}; v-x+e^{(\mu)}, w}
  \chi[v \neq w],
  \label{eq:h_int[2]}
\end{equation}
and
\begin{align}
 h_{\rm int}[\Psi_{\rho, y}, \Psi_{\mu, x, w, w}]
  =
  e^{i k\cdot y}
  \left(
   e^{-i k \cdot v}
   \tilde{U}_{w, y-v+e^{(\rho)}; e^{(\mu)}, x}
   - e^{-i k \cdot w}
   \tilde{U}_{v, y-w+e^{(\rho)}; e^{(\mu)}, x}
  \right),
 \label{eq:h_int[3]}
\end{align}
where the indicator function $\chi$ is defined by $\chi[{\rm true}]=1$
and $\chi[{\rm false}]=0$.

\subsection{Treatment of the Hopping Hamiltonian}

It is easy to calculate matrix elements of hopping Hamiltonian, since the
modified hopping Hamiltonian $\tilde{H}_{\rm hop}$ is ``diagonalized''
in terms of localized electron operators (\ref{eq:def_of_a}) and
(\ref{eq:def_of_b}). We can easily obtain the matrix element
(\ref{eq:h_hop}) from
\begin{equation}
 \tilde{H}_{\rm hop} \Psi_{\mu, A}
  =
  t \lambda^2 ( |A \cap \Lambda^\prime|+\chi[\mu \neq 0])
  \Psi_{\mu, A}.
\end{equation}

\subsection{Treatment of the Interaction Hamiltonian}

Here, we calculate the matrix elements.

\paragraph{Calculation of $h_{\rm int}[\Phi, \Psi_{\mu, x}]$}

First, we consider $H_{\rm int} \Psi_{\mu, x}$,
\begin{align}
 H_{\rm int} \Psi_{\mu, x}
 =
 \sum_{w_1, w_2, w_3, w_4 \in \Lambda} \sum_{v \in \Lambda_o}
 e^{i k \cdot v} \tilde{U}_{w_1, w_2; w_3, w_4}
 a_{w_1, \uparrow}^\dagger a_{w_2, \downarrow}^\dagger
 b_{w_3, \downarrow} b_{w_4, \uparrow}
 a_{v+e^{(\mu)}, \downarrow}^\dagger b_{x+v, \uparrow}
 & \Phi_\uparrow
 \nonumber \\
 =
 \sum_{w_2 \in \Lambda} \sum_{v, w_4 \in \Lambda_o}
 e^{i k \cdot v}
 \Biggl[
 \tilde{U}_{w_4, w_2; v+e^{(\mu)}, w_4}
 a_{w_2, \downarrow}^\dagger b_{x+v, \uparrow}
 - 
 \tilde{U}_{x+v, w_2; v+e^{(\mu)}, w_4}
 a_{w_2, \downarrow}^\dagger b_{w_4, \uparrow}
 \Biggr]
 \chi[w_4 \neq x+v] & \Phi_\uparrow
 +
 \nonumber \\
 \sum_{w_1 \in \Lambda^\prime}
 \sum_{w_2 \in \Lambda} \sum_{v, w_4 \in \Lambda_o}
 e^{i k \cdot v} \tilde{U}_{w_1, w_2; v+e^{(\mu)}, w_4}
 a_{w_2, \downarrow}^\dagger
 b_{w_4, \uparrow} b_{x+v, \uparrow}
 a_{w_1, \uparrow}^\dagger
 \chi[w_4 \neq x+v] & \Phi_\uparrow.
 \label{eq:H_int_psi_mu-x:1}
\end{align}
Here, we decompose $w_2$ into $w \in \Lambda_o$ and $e^{(\rho)}$. We make
the change of variables $w_4=w+y$ and $v=w-x+v$ for first line. Then we
obtain
\begin{align}
 & \sum_{\rho=0}^{d} \sum_{v, w, y \in \Lambda_o}
 \Biggl[
 e^{i k \cdot (v - x+w)}
 \tilde{U}_{w+y, w+e^{(\rho)}; v -x+w+e^{(\mu)}, w+y}
 a_{w+e^{(\rho)}, \downarrow}^\dagger b_{v+w, \uparrow}
 -
 \nonumber \\
 & \qquad
 e^{i k \cdot (w+v-x)}
 \tilde{U}_{v+w, w+e^{(\rho)}; v-x+w+e^{(\mu)}, w+y}
 a_{w+e^{(\rho)}, \downarrow}^\dagger b_{w+y, \uparrow}
 \Biggr]
 \chi[w+y \neq v+w] \Phi_\uparrow
 \nonumber \\
 & =
 \sum_{\rho=0}^{d} \sum_{v, y \in \Lambda_o}
 \chi[y \neq v]
 \biggl[
 e^{i k \cdot (v - x)}
 \tilde{U}_{y, e^{(\rho)}; v-x+e^{(\mu)}, y}
 \Psi_{\rho, v}
 -
 e^{i k \cdot (v-x)}
 \tilde{U}_{v, e^{(\rho)}; v-x+e^{(\mu)}, y}
 \Psi_{\rho, y}
 \biggr] 
 \nonumber \\
 & =
 \sum_{\rho=0}^{d} \sum_{v, y \in \Lambda_o}
 \chi[y \neq v]
 \biggl[
 e^{i k \cdot (y-x)}
 \tilde{U}_{v, e^{(\rho)}; y-x+e^{(\mu)}, v}
 - 
 e^{i k \cdot (v-x)}
 \tilde{U}_{v, e^{(\rho)}; v-x+e^{(\mu)}, y}
 \biggr] \Psi_{\rho, y},
\end{align}
where we used the relation
$\tilde{U}_{x + u, y + u; v + u, w + u} = \tilde{U}_{x, y; v, w}$ with
$u \in \Lambda_o$. Thus we find the representation
(\ref{eq:h_int[1]}). \qed

For second line of (\ref{eq:H_int_psi_mu-x:1}), we make the change of
variables $w_2=w^\prime+e^{(\rho)}$, $w_1=w^\prime+y$, $w_4=w^\prime+w$
and $v=w^\prime-x+v$. Then we obtain
\begin{align}
 & \sum_{\rho = 0}^{d} \sum_{y \in \Lambda^\prime}
 \sum_{v, w^\prime, w \in \Lambda_o}
 e^{i k \cdot (v+w^\prime - x)}
 \tilde{U}_{y+w^\prime, w^\prime+e^{(\rho)}; v+w^\prime - x+e^{(\mu)}, w^\prime+w}
 \times
 \nonumber \\
 & \qquad 
 a_{w^\prime+e^{(\rho)}, \downarrow}^\dagger
 b_{w^\prime+w, \uparrow} b_{v+w^\prime, \uparrow}
 a_{y+w^\prime, \uparrow}^\dagger
 \chi[w^\prime+w \neq v+w^\prime] \Phi_\uparrow
 \nonumber \\
 & = \sum_{\rho = 0}^{d} \sum_{y \in \Lambda^\prime}
 \sum_{v, w \in \Lambda_o}
 e^{i k \cdot (v-x)}
 \tilde{U}_{y, e^{(\rho)}; v-x+e^{(\mu)}, w}
 \chi[v \neq w]
 \sum_{w^\prime \in \Lambda_o}
 e^{i k \cdot w^\prime} T_{w^\prime}
 a_{e^{(\rho)}, \downarrow}^\dagger
 b_{w, \uparrow} b_{v, \uparrow}
 a_{y, \uparrow}^\dagger
 \Phi_\uparrow
 \nonumber \\
 & = \sum_{\rho = 0}^{d} \sum_{w \in \Lambda^\prime}
 \sum_{v, y \in \Lambda_o}
 e^{i k \cdot (v - x)}
 \tilde{U}_{y, e^{(\rho)}; v - x+e^{(\mu)}, w}
 \chi[y \neq v] \Psi_{\rho, y, v, w}.
\end{align}
Thus we find representation (\ref{eq:h_int[2]}). \qed

\paragraph{Calculation of $h[\Psi_{\rho, y}, \Psi_{\mu, x, v, w}]$}

Next, we consider $H_{\rm int}\Psi_{\mu, x, v, w}$,
\begin{align}
 & H_{\rm int} \Psi_{\mu, x, v, w}
 = \sum_{w_1, w_2, w_3, w_4 \in \Lambda} \sum_{w_5 \in \Lambda_o}
 e^{i k\cdot w_5} \tilde{U}_{w_1, w_2; w_3, w_4}
 \times
 \nonumber \\
 & \qquad
 a_{w_2, \downarrow}^\dagger b_{w_3, \downarrow}
 a_{w_5+e^{(\mu)}, \downarrow}^\dagger
 a_{w_1, \uparrow}^\dagger
 b_{w_5+w, \uparrow} b_{w_5+v, \uparrow}
 b_{w_4, \uparrow} a_{w_5+x, \uparrow}^\dagger
 \Phi_\uparrow
 \nonumber \\
 & = \sum_{\rho=0}^d \sum_{w_1 \in \Lambda} \sum_{w_2, w_5 \in \Lambda_o}
 e^{i k\cdot w_5} \tilde{U}_{w_1, w_2+e^{(\rho)}; w_5+e^{(\mu)}, w_5+x}
 \times
 \nonumber \\
 & \qquad
 a_{w_2+e^{(\rho)}, \downarrow}^\dagger
 a_{w_1, \uparrow}^\dagger
 b_{w_5+w, \uparrow} b_{w_5+v, \uparrow}
 \Phi_\uparrow
 + \mbox{other terms}
 \nonumber \\
 & = \sum_{\rho=0}^d \sum_{w_2, w_5 \in \Lambda_o}
 e^{i k\cdot w_5}  a_{w_2+e^{(\rho)}, \downarrow}^\dagger
 \biggl(
 \tilde{U}_{w, w_2+e^{(\rho)}-w_5; e^{(\mu)}, x}
 b_{w_5+v, \uparrow}
 -
 \nonumber \\
 & \qquad
 \tilde{U}_{v, w_2+e^{(\rho)}-w_5; e^{(\mu)}, x}
 b_{w_5+w, \uparrow}
 \biggr)
 \Phi_\uparrow
 + \mbox{other terms}
 \nonumber \\
 & = \sum_{\rho=0}^d \sum_{w_2, w_5 \in \Lambda_o}
 e^{i k\cdot (w_5+w_2)}	 a_{w_2+e^{(\rho)}, \downarrow}^\dagger
 \biggl(
 \tilde{U}_{w, w_5+e^{(\rho)}; e^{(\mu)}, x}
 b_{w_2+w_5+v, \uparrow}
 -
 \nonumber \\
 & \qquad
 \tilde{U}_{v, w_5+e^{(\rho)}; e^{(\mu)}, x}
 b_{w_2+w_5+w, \uparrow}
 \biggr)
 \Phi_\uparrow
 + \mbox{other terms}
 \nonumber \\
 & = \sum_{\rho=0}^d \sum_{w_5 \in \Lambda_o}
 e^{i k\cdot w_5}
 \biggl(
 \tilde{U}_{w, w_5+e^{(\rho)}; e^{(\mu)}, x}
 \Psi_{\rho, w_5+v}
 -
 \nonumber \\
 & \qquad \tilde{U}_{v, w_5+e^{(\rho)}; e^{(\mu)}, x}
 \Psi_{\rho, w_5+w}
 \biggr)
 + \mbox{other terms}
 \nonumber \\
 & = \sum_{\rho=0}^d \sum_{y \in \Lambda_o}
 e^{i k\cdot y}
 \biggl(
 e^{-i k \cdot v}
 \tilde{U}_{w, y-v+e^{(\rho)}; e^{(\mu)}, x}
 -
 \nonumber \\
 & \qquad
 e^{-i k \cdot w}
 \tilde{U}_{v, y-w+e^{(\rho)}; e^{(\mu)}, x}
 \biggr)
 \Psi_{\rho, y}
 + \mbox{other terms},
\end{align}
where ``other terms'' do not contribute to the matrix elements
$h_{\rm int}[\Psi_{\rho, y}, \Psi_{\mu, x, v, w}]$. Thus we find the
representation (\ref{eq:h_int[3]}). \qed

\section{Estimates of the Matrix Elements \label{sec:est_ME}}

In this section, we show the estimates of matrix elements and to prove
Lemma \ref{lemma:D[Psi]}. Before showing the details of estimates, we
list the results of estimates and give the proof of Lemma
\ref{lemma:D[Psi]}.

\subsection{Summary of Estimates}

Here, we obtain following bounds:
\begin{align}
 \Re[h_{\rm int}[\Omega, \Omega]] \geq &
 \frac{2 U}{\lambda^4}
 \biggl[
 \frac{d (|q|+|q|^{-1})}{2}-
 \sum_{j=1}^d\cos(k \cdot 2e^{(j)} + \theta)
 -
 \frac{G_1}{\lambda^2}
 \biggr],
 \label{eq:bound_h_int[1]}
\end{align}
\begin{equation}
 \Re[h_{\rm int}[\Psi_{0, x}, \Psi_{0, x}]] \geq
  U\left( \chi[x \neq o] - \frac{G_2}{\lambda^2} \right),
  \label{eq:bound_h_int[2]}
\end{equation}
\begin{equation}
 |h_{\rm int}[\Psi_{0, x}, \Omega]|
  \leq
  \frac{U}{\alpha} \frac{G_3}{\lambda^4},
  \label{eq:bound_h_int[3]}
\end{equation}
\begin{equation}
  |h_{\rm int}[\Psi_{\mu, y, v, w}, \Omega]|
  \leq \frac{U}{\alpha} \frac{G_4}{\lambda^2},
  \label{eq:bound_h_int[4]}
\end{equation}
\begin{equation}
 \sum_{\rho=0}^d \sum_{v_1 \in \Lambda_o}
  \chi[(\rho, v_1) \neq (0, x)]
  |h_{\rm int}[\Psi_{0, x}, \Psi_{\rho, v_1}]|
  \leq
  U \frac{G_5}{\lambda},
  \label{eq:bound_h_int[5]}
\end{equation}
\begin{equation}
 \sum_{\rho=0}^d \sum_{y_1 \in \Lambda^\prime}
  \sum_{v_1, w_1 \in \Lambda_o}
  \chi[v_1 \neq w_1]
  |h_{\rm int}[\Psi_{0, x}, \Psi_{\rho, y_1, v_1, w_1}]|
  \leq
  U \frac{G_6}{\lambda},
  \label{eq:bound_h_int[6]}
\end{equation}
and
\begin{align}
 & \Re[h_{\rm int}[\Psi_{\mu, A}, \Psi_{\mu, A}]]
 -
 \nonumber \\
 & \sum_{\rho=0}^d \sum_{B \subset \Lambda}
 \chi[(\rho, B) \neq (\mu, A)]
 |h_{\rm int}[\Psi_{\mu, A}, \Psi_{\rho, B}]|
 \geq
 U \left[ \chi[e^{(\mu)} \in A] - \frac{G_7}{\lambda} \right],
 \label{eq:bound_h_int[7]}
\end{align}
where $x, v, w \in \Lambda_o$ with
$v \neq w$, $y \in \Lambda^\prime$ and $\mu = 0, 1, \cdots, d$.
$G_{l}$ $(l=1, 2, \cdots, 7)$ are finite constants which do not depend
on the system size.

\subsection{Proof of Lemma \ref{lemma:D[Psi]} \label{sec:estimates_of_D[Psi]}}

Here, we show the estimates of each $D[\Psi_{\mu, A}]$. We start to
evaluate cases that $\mu=0$ and $A\subset \Lambda_o$.

\paragraph{Estimates of $D[\Omega]$}

Since $\tilde{H}_{\rm hop} \Omega=0$, we can represent
\begin{align}
 D[\Omega] = &\Re [h_{\rm int}[\Omega, \Omega]]
  -\sum_{\Psi \in \B_k\backslash\{\Omega\}} |h_{\rm int}[\Omega, \Psi]|
 \nonumber \\
 = & \Re[h_{\rm int}[\Omega, \Omega]]
 - \sum_{\rho=0}^d \sum_{v_1 \in \Lambda_o}
 \chi[(\rho, v_1) \neq (0, o)]
 |h_{\rm int}[\Omega, \Psi_{\rho, v_1}]|
 - \nonumber \\
 & \qquad \sum_{\rho=0}^d \sum_{y_1 \in \Lambda^\prime}
 \sum_{v_1, w_1 \in \Lambda_o}
 \chi[v_1 \neq w_1]
 |h_{\rm int}[\Omega, \Psi_{\rho, y_1, v_1, w_1}]|.
\end{align}
Note that $\Omega = \Psi_{0, o}/\alpha$ and $\alpha > 0$.
Then we obtain (\ref{eq:D[Omega]}), from eqs. (\ref{eq:bound_h_int[1]}),
(\ref{eq:bound_h_int[5]}) and (\ref{eq:bound_h_int[6]}). \qed

\paragraph{Estimates of $D[\Psi_{0, x}]$}
$\tilde{H}_{\rm hop} \Psi_{0, x}$ also vanishes, then we have
\begin{align}
 D[\Psi_{0, x}] = & \Re[h_{\rm int}[\Psi_{0, x}, \Psi_{0, x}]]
 - \sum_{\rho=0}^d \sum_{v_1 \in \Lambda_o}
 \chi[(\rho, v_1) \neq (0, x)]
 \chi[(\rho, v_1) \neq (0, o)]
 |h_{\rm int}[\Psi_{0, x}, \Psi_{\rho, v_1}]|
 - \nonumber \\
 & \qquad
 |h_{\rm int}[\Psi_{0, x}, \Omega]|
 -
 \sum_{\rho=0}^d \sum_{y_1 \in \Lambda^\prime}
 \sum_{v_1, w_1 \in \Lambda_o}
 \chi[v_1 \neq w_1]
 |h_{\rm int}[\Psi_{0, x}, \Psi_{\rho, y_1, v_1, w_1}]|.
\end{align}
Thus we obtain (\ref{eq:D[Psi_0x]}) from (\ref{eq:bound_h_int[2]}),
(\ref{eq:bound_h_int[3]}), (\ref{eq:bound_h_int[5]}) and (\ref{eq:bound_h_int[6]}).
\qed

\paragraph{Estimates of Other Contributions}
We estimate $D[\Psi_{\mu, A}]$ with $\mu\neq 0$,
$A \cap \Lambda^\prime \neq \emptyset$ or both.
Here, we regard $\Psi_{\mu, x}$ and $\Psi_{\mu, y, v, w}$ as
$\Psi_{\mu, \Lambda_o\backslash\{x\}}$ and
$\Psi_{\mu, (\Lambda_o\backslash\{v, w\})\cup\{y\}}$
since the differences between them are only sign factor which is
irrelevant to absolute value of off-diagonal matrix elements.
Thus we have
\begin{align}
 D[\Psi_{\mu, A}] = &
 \Re[h_{\rm hop}[\Psi_{\mu, A}, \Psi_{\mu, A}]+
 h_{\rm int}[\Psi_{\mu, A}, \Psi_{\mu, A}]]-
 \nonumber \\
 &
 \sum_{\rho =0}^d \sum_{B \subset \Lambda}
 \chi[(\rho, B)\neq(0, \Lambda_o\backslash\{o\}),(\mu, A)]
 |h_{\rm int}[\Psi_{\mu, A}, \Psi_{\rho, B}]|
 -|h_{\rm int}[\Psi_{\mu, A}, \Omega]|.
\end{align}
Since only $\Psi_{\mu, y, v, w}$ has non-vanishing
off-diagonal matrix elements related $\Omega$:
$h_{\rm int}[\Psi_{\mu, A}, \Omega]$ with
$A \cap \Lambda^\prime \neq \emptyset$, we have
\begin{equation}
 h_{\rm int}[\Psi_{\mu, A}, \Omega] < \frac{U}{\alpha} \frac{G_4}{\lambda^2}
\end{equation}
from (\ref{eq:bound_h_int[4]}). Then we have
\begin{align}
 D[\Psi_{\mu, A}] \geq & t\lambda^2
  + U \left[ \chi[e^{(\mu)} \in A] - \frac{G_7}{\lambda} \right]
  - \frac{U}{\alpha} \frac{G_4}{\lambda^2}
 \nonumber \\
 & \geq t\lambda^2 - \frac{U}{\lambda}G_7
 - \frac{U}{\alpha} \frac{G_4}{\lambda^2}
\end{align}
from (\ref{eq:h_hop}) and (\ref{eq:bound_h_int[7]}), where we use
\begin{equation}
 h_{\rm hop}[\Psi_{\mu, A}, \Psi_{\mu, A}] =
  t \lambda^2 (\chi[\mu\neq 0] + |A \cap \Lambda^\prime|) \geq t \lambda^2
\end{equation}
for the cases $\mu\neq 0$, $A\cap\Lambda^\prime \neq \emptyset$ or both. 
Thus, we find the bounds for all contributions and we conclude
Lemma \ref{lemma:D[Psi]}. \qed

\subsection{Estimates of Matrix Elements}

Here, we estimate the matrix elements.
To estimate the matrix elements, it is convenient to introduce
$\varphi_{x, \sigma}^{(y)}$ by
$\varphi_{x, \sigma}^{(y)} = \tilde{\psi}_{x, \sigma}^{(y)} - \psi_{x, \sigma}^{(y)}$.
Note that $\varphi_{x, \sigma}^{(y)} = \tilde{\psi}_{x, \sigma}^{(y)}$
for $|x - y| \geq 1$. Lemma \ref{lemma:bound_sum_psi} and
\ref{lemma:bound_psi} give us the bound of
$\sum_{x}|\varphi_{x, \sigma}^{(y)}|$ and $|\varphi_{x, \sigma}^{(y)}|$.
Then, we obtain bounds of the matrix elements from representations of
each matrix element in terms of $\varphi_{x, \sigma}^{(y)}$.

\paragraph{Estimates of $\Re [h_{\rm int}[\Psi_{0, x}, \Psi_{0, x}]]$}

First, we estimate the diagonal part of the matrix elements.
From the representation (\ref{eq:h_int[1]}), we can write
\begin{align}
 & \frac{h_{\rm int} [\Psi_{0, x}, \Psi_{0, x}]}{U}
 =
 \frac{1}{U} \sum_{v \in \Lambda_o}
 \left(
 \tilde{U}_{v, o; o, v}
 - e^{i k \cdot v}
 \tilde{U}_{x+v, o; v, x}
 \right)
 \nonumber \\
 & =
 \sum_{v \in \Lambda_o} \sum_{w \in \Lambda}
 \Biggl[
 (\tilde{\psi}_{w, \uparrow}^{(v)} \tilde{\psi}_{w, \downarrow}^{(o)})^\ast
 \psi_{w, \downarrow}^{(o)} \psi_{w, \uparrow}^{(v)}
 -
 e^{i k \cdot v}
 (\tilde{\psi}_{w, \uparrow}^{(x+v)} \tilde{\psi}_{w, \downarrow}^{(o)})^\ast
 \psi_{w, \downarrow}^{(v)} \psi_{w, \uparrow}^{(x)}
 \Biggr].
\end{align}
From the explicit representation of $\psi_{x, \sigma}^{(y)}$, we obtain
\begin{align}
 & \frac{h_{\rm int} [\Psi_{0, x}, \Psi_{0, x}]}{U}
 =
 (\tilde{\psi}_{o, \uparrow}^{(o)} \tilde{\psi}_{o, \downarrow}^{(o)})^\ast
 -
 e^{i k \cdot x}
 (\tilde{\psi}_{-x, \uparrow}^{(o)} \tilde{\psi}_{x,
 \downarrow}^{(o)})^\ast
 +
 \nonumber \\
 & \quad 
 \frac{1}{\lambda^2} \sum_{j=1}^d
 \left[
 \tilde{\psi}_{-e^{(j)}, \uparrow}^{(o)} \tilde{\psi}_{-e^{(j)}, \downarrow}^{(o)}
 +\tilde{\psi}_{e^{(j)}, \uparrow}^{(o)} \tilde{\psi}_{e^{(j)}, \downarrow}^{(o)}
 \right]^\ast 
 -
 \nonumber \\
 & \quad
 \frac{e^{i k \cdot x}}{\lambda^2} \sum_{j=1}^d
 \biggl[
 \tilde{\psi}_{-x-e^{(j)}, \uparrow}^{(o)} \tilde{\psi}_{x-e^{(j)}, \downarrow}^{(o)}
 +
 \tilde{\psi}_{-x+e^{(j)}, \uparrow}^{(o)} \tilde{\psi}_{x+e^{(j)}, \downarrow}^{(o)}
 \biggr]^\ast
 +
 \nonumber \\
 &
 \frac{1}{\lambda^2} \sum_{j=1}^{d}
 \Biggl\{
 q^{-1/2}
 \left[
 \tilde{\psi}_{e^{(j)}, \uparrow}^{(o)} \tilde{\psi}_{-e^{(j)}, \downarrow}^{(o)}
 -
 e^{-i k \cdot (x+2e^{(j)})}
 \tilde{\psi}_{-x-e^{(j)}, \uparrow}^{(o)} \tilde{\psi}_{x+e^{(j)}, \downarrow}^{(o)}
 \right]^\ast +
 \nonumber \\
 & \quad
 q^{1/2}
 \left[
 \tilde{\psi}_{-e^{(j)}, \uparrow}^{(o)} \tilde{\psi}_{e^{(j)}, \downarrow}^{(o)}
 -
 e^{-i k \cdot (x-2e^{(j)})}
 \tilde{\psi}_{-x+e^{(j)}, \uparrow}^{(o)} \tilde{\psi}_{x-e^{(j)}, \downarrow}^{(o)}
 \right]^\ast
 \Biggr\},
 \label{eq:h_int[Psi_0xPsi_0x]}
\end{align}
where we use the relation
$\tilde{\psi}_{x, \sigma}^{(y+u)} = \tilde{\psi}_{x-u, \sigma}^{(y)}$
for $u \in \Lambda_o$.

For $x = o$, we obtain
\begin{align}
 & \frac{h[\Psi_{0, o}, \Psi_{0, o}]}{U}
 =
 \frac{d (|q|+|q|^{-1})-2\sum_{j=1}^d\cos(k \cdot 2e^{(j)} + \theta)}{\lambda^4}
 +
 \nonumber \\
 &
 \frac{1}{\lambda^2}
 \sum_{j=1}^d
 \Biggl\{
 \left[
 \frac{q^{-1/4}}{\lambda}
 (\varphi_{e^{(j), \uparrow}}^{(o)}+ \varphi_{-e^{(j)}, \downarrow}^{(o)})
 +
 \varphi_{e^{(j)}, \uparrow}^{(o)} \varphi_{-e^{(j)}, \downarrow}^{(o)}
 \right]^\ast \times
 \nonumber \\
 & \qquad
 (q^{-1/2} - q^{1/2} e^{-i k \cdot 2e^{(j)}})
 +
 \nonumber \\
 &
 \left[
 \frac{q^{1/4}}{\lambda}
 (\varphi_{e^{(j), \downarrow}}^{(o)}+ \varphi_{-e^{(j)}, \uparrow}^{(o)})
 +\frac{1}{\lambda^2}
 \varphi_{e^{(j)}, \downarrow}^{(o)} \varphi_{-e^{(j)}, \uparrow}^{(o)}
 \right]^\ast \times
 \nonumber \\
 & \qquad
 (q^{1/2} - q^{-1/2} e^{i k \cdot 2e^{(j)}})
 \Biggr\}.
\end{align}
Lemma \ref{lemma:bound_psi} means
$|\varphi_{\pm e^{(j)}, \sigma}^{(o)}|\leq B_3 \lambda^{-3}$, then we
find the bound (\ref{eq:bound_h_int[1]}).\qed

For $x \neq o$, we find (\ref{eq:bound_h_int[2]}) from
$|\tilde{\psi}_{o, \sigma}^{(o)}| \leq 1-O(\lambda^{-2})$ and
$|\tilde{\psi}_{x, \sigma}^{(o)}| = O(\lambda^{-2})$ for $x \neq o$. \qed

\paragraph{Estimates of $h_{\rm int}[\Psi_{0, y}, \Psi_{0, o}]$}

Next, we estimate the off-diagonal part of the matrix elements.
To obtain (\ref{eq:bound_h_int[3]}), we calculate the following matrix element
\begin{align}
 h_{\rm int} & [\Psi_{0, y}, \Psi_{0, o}] =
 \sum_{v \in \Lambda_o}
 \left[
 e^{i k \cdot y} \tilde{U}_{v, o; y, v}
 - e^{i k \cdot v} \tilde{U}_{v, o; v, y}
 \right]
 \nonumber \\
 = &
 \frac{U}{\lambda^2} e^{i k \cdot y}
 \sum_{j=1}^{d}
 \Biggl[
 (q^{-1/2} - q^{1/2} e^{-2i k \cdot e^{(j)}})
 (\tilde{\psi}_{e^{(j)}, \uparrow}^{(o)} \tilde{\psi}_{y-e^{(j)}, \downarrow}^{(o)})^\ast
 + \nonumber \\
 & \quad
 (q^{1/2} - q^{-1/2} e^{2i k \cdot e^{(j)}})
 (\tilde{\psi}_{-e^{(j)}, \uparrow}^{(o)} \tilde{\psi}_{y+e^{(j)}, \downarrow}^{(o)})^\ast
 \Biggr].
\end{align}
Since $|\tilde{\psi}_{w}^{(o)}| = O(\lambda^{-1})$ for
$w \in \Lambda^\prime$, we find the bound (\ref{eq:bound_h_int[3]}).

\paragraph{Estimates of $h[\Psi_{\mu, x, v, w}, \Psi_{0, o}]$}
Using the following bound
\begin{align}
 |h[\Psi_{\mu, y, v, w}, \Psi_{0, o}]|
 = &
 \frac{U}{\lambda^2}
 \biggl|
 \sum_{j=1}^d
 \biggl[
 q^{-1/2} \delta_{v, w+2e^{(j)}}
 (\tilde{\psi}_{w+e^{(j)}}^{(o)} \tilde{\psi}_{w+e^{(j)}}^{(e^{(\mu)})})
 + \nonumber \\
 & \qquad
 q^{1/2} \delta_{v, w-2e^{(j)}}
 (\tilde{\psi}_{w-e^{(j)}}^{(o)} \tilde{\psi}_{w-e^{(j)}}^{(e^{(\mu)})})
 \biggr]
 \biggr|
 \nonumber \\
 & \leq O(U \lambda^{-2}),
\end{align}
we obtain (\ref{eq:bound_h_int[4]}).

\paragraph{Estimates of $h[\Psi_{0, x}, \Psi_{\mu, y}]$ with $x \neq o$}
Using the following bound
\begin{align}
 & \sum_{\rho=0}^d \sum_{y \in \Lambda_o}
 |h_{\rm int}[\Psi_{0, x}, \Psi_{\rho, y}]|
 \chi[(\rho, y)\neq(0, x)]
 \nonumber \\
 & =
 U \sum_{\rho=0}^d \sum_{y \in \Lambda_o}
 \left|
 \sum_{v, w \in \Lambda}
 (\tilde{\psi}_{w, \uparrow}^{(v)}
 \tilde{\psi}_{w, \downarrow}^{(e^{(\rho)})})^\ast
 (\delta_{v, x-y} \delta_{w, x-y}
 -e^{i k \cdot v} \delta_{v, x+y} \delta_{w, x})
 \right| \times
 \nonumber \\
 & \qquad
 \chi[(\rho, y)\neq(0, x)]
 + O(U \lambda^{-2})
 \nonumber \\
 & =
 U \sum_{\rho=0}^d \sum_{y \in \Lambda_o}
 \left|
 \tilde{\psi}_{o, \uparrow}^{(o)}
 \tilde{\psi}_{x-y, \downarrow}^{(e^{(\rho)})}
 -
 e^{-i k \cdot (x+y)}
 \tilde{\psi}_{x, \uparrow}^{(x+y)}
 \tilde{\psi}_{x, \downarrow}^{(e^{(\rho)})}
 \right|
 \chi[(\rho, y)\neq(0, x)]
 + O(U \lambda^{-2})
 \nonumber \\
 & =
 U \sum_{y \in \Lambda_o}
 \left|
 \tilde{\psi}_{o, \uparrow}^{(o)}
 \tilde{\psi}_{x-y, \downarrow}^{(o)}
 -
 e^{-i k \cdot (x+y)}
 \tilde{\psi}_{x, \uparrow}^{(x+y)}
 \tilde{\psi}_{x, \downarrow}^{(o)}
 \right|
 \chi[(\rho, y)\neq(0, x)]
 + O(U \lambda^{-1})
 \nonumber \\
 & =
 U \sum_{y \in \Lambda_o}
 \left|
 \tilde{\psi}_{o, \uparrow}^{(o)}
 \tilde{\psi}_{x-y, \downarrow}^{(o)}
 -
 e^{-i k \cdot (x+y)}
 \tilde{\psi}_{x, \uparrow}^{(x+y)}
 \tilde{\psi}_{x, \downarrow}^{(o)}
 \right|
 - 
 \left|
 \tilde{\psi}_{o, \uparrow}^{(o)}
 \tilde{\psi}_{o, \downarrow}^{(o)}
 -
 e^{-2 i k \cdot x}
 \tilde{\psi}_{x, \uparrow}^{(2x)}
 \tilde{\psi}_{x, \downarrow}^{(o)}
 \right|
 + O(U \lambda^{-1})
 \nonumber \\
 & \leq
 U \sum_{y \in \Lambda_o}
 \left(
 \left|
 \tilde{\psi}_{o, \uparrow}^{(o)}
 \tilde{\psi}_{x-y, \downarrow}^{(o)}
 \right|
 +
 \left|
 \tilde{\psi}_{x, \uparrow}^{(x+y)}
 \tilde{\psi}_{x, \downarrow}^{(o)}
 \right|
 \right)
 - 
 \left|
 \tilde{\psi}_{o, \uparrow}^{(o)}
 \tilde{\psi}_{o, \downarrow}^{(o)}
 \right|
 +
 \left|
 \tilde{\psi}_{x, \uparrow}^{(2x)}
 \tilde{\psi}_{x, \downarrow}^{(o)}
 \right|
 + O(U \lambda^{-1})
 \nonumber \\
 &
 \leq O(U \lambda^{-1}).
\end{align}
for $x \neq o$, and using another bound
\begin{align}
 & \sum_{\rho=0}^d \sum_{y \in \Lambda_o}
 |h_{\rm int}[\Psi_{0, o}, \Psi_{\rho, y}]|
 \chi[(\rho, y)\neq(0, o)]
 \nonumber \\
 & =
 U \sum_{y \in \Lambda_o}
 \left|
 \tilde{\psi}_{o, \uparrow}^{(o)}
 \tilde{\psi}_{-y, \downarrow}^{(o)}
 -
 e^{-i k \cdot y}
 \tilde{\psi}_{o, \uparrow}^{(y)}
 \tilde{\psi}_{o, \downarrow}^{(o)}
 \right|
 + O(\lambda^{-1})
 \nonumber \\
 & \leq
 U \sum_{y \in \Lambda_o}
 \left(
 \left|
 \tilde{\psi}_{o, \uparrow}^{(o)}
 \tilde{\psi}_{-y, \downarrow}^{(o)}
 \right|
 +
 \left|
 \tilde{\psi}_{o, \uparrow}^{(y)}
 \tilde{\psi}_{o, \downarrow}^{(o)}
 \right|
 \right)
 \chi[y \neq o]
 + O(U \lambda^{-1})
 \nonumber \\
 & =
 U \sum_{y \in \Lambda_o}
 \left(
 \left|
 \tilde{\psi}_{o, \uparrow}^{(o)}
 \tilde{\psi}_{-y, \downarrow}^{(o)}
 \right|
 +
 \left|
 \tilde{\psi}_{o, \uparrow}^{(y)}
 \tilde{\psi}_{o, \downarrow}^{(o)}
 \right|
 \right)
 - 2 \left|
 \tilde{\psi}_{o, \uparrow}^{(o)}
 \tilde{\psi}_{o, \downarrow}^{(o)}
 \right|
 + O(U \lambda^{-1})
 \nonumber \\
 &
 \leq O(U \lambda^{-1}),
\end{align}
for $x=o$, we obtain (\ref{eq:bound_h_int[5]}).

\paragraph{Estimates of $h[\Psi_{0, x}, \Psi_{\mu, y, v, w}]$}
Using the following bound
\begin{align}
 & \sum_{\mu=0}^d \sum_{y \in \Lambda^\prime}
 \sum_{v, w \in \Lambda_o}
 |h_{\rm int}[\Psi_{0, x}, \Psi_{\mu, y, v, w}]|
 \nonumber \\
 & \leq
 U \sum_{j=1}^{d}
 \sum_{y \in \Lambda_j} \sum_{v, w \in \Lambda_o}
 \left|
 e^{i k \cdot v} \tilde{\psi}_{e^{(j)}, \uparrow}^{(w)}
 \tilde{\psi}_{e^{(j)}, \downarrow}^{(x-v)}
 +
 e^{i k \cdot w} \tilde{\psi}_{e^{(j)}, \uparrow}^{(v)}
 \tilde{\psi}_{e^{(j)}, \downarrow}^{(x-w)}
 \right| +
 O(U\lambda^{-1})
 \nonumber \\
 & \leq
 U \sum_{j=1}^{d}
 \sum_{y \in \Lambda_j} \sum_{v, w \in \Lambda_o}
 \left(
 |\tilde{\psi}_{e^{(j)}, \uparrow}^{(w)}|
 |\tilde{\psi}_{e^{(j)}, \downarrow}^{(x-v)}|
 +
 |\tilde{\psi}_{e^{(j)}, \uparrow}^{(v)}|
 |\tilde{\psi}_{e^{(j)}, \downarrow}^{(x-w)}|
 \right)
 + O(U\lambda^{-1})
 \nonumber \\
 & \leq O(U\lambda^{-1}),
\end{align}
we obtain  (\ref{eq:bound_h_int[6]}).

\paragraph{Estimates of Other Matrix Elements}

To obtain  (\ref{eq:bound_h_int[7]}), we calculate the
matrix elements of $H_{\rm int}$.
\begin{align}
 & H_{\rm int} \Psi_{\mu, A}
 =
 \sum_{w_1, w_2, w_3, w_4 \in \Lambda} \tilde{U}_{w_1, w_2; w_3, w_4}
 a_{w_1, \uparrow}^\dagger a_{w_2, \downarrow}^\dagger
 b_{w_3, \downarrow} b_{w_4, \uparrow}
 \times
 \nonumber \\
 & \qquad 
 \sum_{w \in \Lambda_o} e^{i k \cdot w}
 a_{w+e^{(\mu)}, \downarrow}^\dagger
 \left(
 \prod_{v \in A} a_{v+w, \uparrow}^\dagger
 \right)
 \Psi_{\rm vac}
 \nonumber \\
 & =
 \sum_{w \in \Lambda_o} \sum_{w_1, w_2, w_3, w_4 \in \Lambda}
 e^{i k \cdot w}
 \tilde{U}_{w_1, w_2; w_3, w_4}
 a_{w_2, \downarrow}^\dagger
 b_{w_3, \downarrow}
 a_{w+e^{(\mu)}, \downarrow}^\dagger
 a_{w_1, \uparrow}^\dagger
 b_{w_4, \uparrow}
 \left(
 \prod_{v \in A} a_{v+w, \uparrow}^\dagger
 \right)
 \Psi_{\rm vac}
 \nonumber \\
 & =
 \sum_{w \in \Lambda_o} \sum_{w_1, w_2, w_4 \in \Lambda}
 e^{i k \cdot w}
 \tilde{U}_{w_1, w_2; w+e^{(\mu)}, w_4}
 \chi[w_4 \in A+w, w_1 \notin (A+w)\backslash \{ w_4 \}] \times
 \nonumber \\
 & \qquad {\rm sgn}[A+w, w_1, w_4]
 a_{w_2, \downarrow}^\dagger
 \left(
 \prod_{v \in ((A+w)\backslash \{ w_4 \}) \cup \{ w_1 \} } a_{v, \uparrow}^\dagger
 \right)
 \Psi_{\rm vac}
 \nonumber
\end{align}
Here, we make the change of the variables $w_2 \rightarrow y+e^{(\rho)}$,
$w \rightarrow w+y$, $w_1 \rightarrow w_1+y$ and
$w_4 \rightarrow w_4+y$. Then we obtain
\begin{align}
 = &
 \sum_{\rho = 0}^{d} \sum_{w, y \in \Lambda_o} \sum_{w_1, w_4 \in \Lambda}
 e^{i k \cdot (w+y)}
 \tilde{U}_{w_1+y, e^{(\rho)}+y; w+y+e^{(\mu)}, w_4+y} \times
 \nonumber \\
 & \quad \chi[w_4+y \in A+w+y,
 w_1+y \notin (A+w+y)\backslash \{ w_4+y \}] \times
 \nonumber \\
 & \quad {\rm sgn}[A+w+y, w_1+y, w_4+y] \times
 \nonumber \\
 & \quad
 a_{e^{(\rho)+y}, \downarrow}^\dagger
 \left(
 \prod_{v \in ((A+w)\backslash \{ w_4 \}) \cup \{ w_1 \}+y} a_{v, \uparrow}^\dagger
 \right)
 \Psi_{\rm vac}
 \nonumber \\
 = &
 \sum_{\rho = 0}^{d} \sum_{w \in \Lambda_o} \sum_{w_1, w_4 \in \Lambda}
 e^{i k \cdot w}
 \tilde{U}_{w_1, e^{(\rho)}; w+e^{(\mu)}, w_4} \times
 \nonumber \\
 & \quad
 \chi[w_4 \in A+w, w_1 \notin (A+w)\backslash \{ w_4 \}]
 {\rm sgn}[A+w, w_1, w_4] \times
 \nonumber \\
 & \quad
 \sum_{y \in \Lambda_o} e^{i k \cdot y}
 a_{e^{(\rho)+y}, \downarrow}^\dagger
 \left(
 \prod_{v \in ((A+w)\backslash \{ w_4 \}) \cup \{ w_1 \}+y} a_{v, \uparrow}^\dagger
 \right)
 \Psi_{\rm vac}
 \nonumber \\
 = &
 \sum_{\rho = 0}^{d} \sum_{w \in \Lambda_o} \sum_{w_1, w_4 \in \Lambda}
 e^{i k \cdot w}
 \tilde{U}_{w_1, e^{(\rho)}; w+e^{(\mu)}, w_4} \times
 \nonumber \\
 & \quad \chi[w_4 \in A+w, w_1 \notin (A+w)\backslash \{ w_4 \}]
 {\rm sgn}[A+w, w_1, w_4] \times
 \nonumber \\
 & \quad
 \Psi_{\rho, ((A+w)\backslash \{ w_4 \}) \cup \{ w_1 \}}.
\end{align}
Thus, we find
\begin{align}
 h_{\rm int}
 [\Psi_{\rho, ((A+w)\backslash \{ w_4 \}) \cup \{ w_1 \}},
 \Psi_{\mu, A}]
 = e^{i k \cdot w}
  \tilde{U}_{w_1, e^{(\rho)}; w+e^{(\mu)}, w_4}
 \chi[w_4 \in A+w, w_1 \notin (A+w)\backslash \{ w_4 \}]
\end{align}
Here, we neglect the sign of the matrix element, 
since it is irrelevant to obtain the bounds. If we
define $A^\prime$ by $A^\prime = ((A+w)\backslash \{ w_4 \}) \cup \{ w_1 \}$, then
we obtain its solution $A=(A^\prime\backslash\{w_1\})\cup\{w_4\}-w$. 
Thus, we obtain
\begin{align}
 & h_{\rm int}
 [\Psi_{\rho, A},
 \Psi_{\mu, (A\backslash\{w_1\})\cup\{w_4\}-w}]
 = e^{i k \cdot w}
 \tilde{U}_{w_1, e^{(\rho)}; w+e^{(\mu)}, w_4}
 \chi[w_1 \in A, w_4 \notin A\backslash \{ w_1 \}].
\end{align}
If $w_1=w_4$ and $w=o$, then $A=(A\backslash\{w_1\})\cup\{w_4\}-w$.
Since other $w_1, w_4$ and $w$ makes
$(A\backslash\{w_1\})\cup\{w_4\}-w=A$, we find
\begin{align}
 & \Re[h_{\rm int}[\Psi_{\rho, A}, \Psi_{\rho, A}]]
 - \sum_{\Phi \in \B\backslash\{\Psi_{\rho, A}\}}
 |h_{\rm int} [\Psi_{\rho, A}, \Phi]|
 \nonumber \\
 & \geq
 \Re\left[ \sum_{w_1 \in A}
 \tilde{U}_{w_1, e^{(\rho)}; e^{(\rho)}, w_1}\right]
 -
 \sum_{w_1 \in A}
 \sum_{w \in \Lambda} \sum_{w_4 \in (\Lambda\backslash A)\cup\{ w_1 \}}
 |\tilde{U}_{w_1, e^{(\rho)}; w, w_4}|
 \chi[(w, w_1)\neq(e^{(\rho)}, w_4)].
 \label{eq:D[general-pre]}
\end{align}
To estimate the first term in (\ref{eq:D[general-pre]}), we use
\begin{align}
 \Re\left[ \sum_{w_1 \in A} \tilde{U}_{w_1, e^{(\rho)}; e^{(\rho)}, w_1} \right]
 \geq & U \Re
 \left[ \sum_{w_1 \in A}
 ( \tilde{\psi}_{e^{(\rho)}, \uparrow}^{(w_1)}
 \tilde{\psi}_{e^{(\rho)}, \downarrow}^{(e^{(\rho)})})^\ast
 \right]
 - O(U \lambda^{-2})
 \nonumber \\
 \geq & U \chi[e^{(\rho)} \in A] - O(U \lambda^{-1}).
 \label{eq:D[general-R]}
\end{align}
We estimate the second term in (\ref{eq:D[general-pre]}). From the
explicit form of $\psi_{x, \sigma}^{(y)}$, we obtain
\begin{align}
 &
 \sum_{w_1 \in A}
 \sum_{w \in \Lambda} \sum_{w_4 \in (\Lambda\backslash A)\cup\{ w_1 \}}
 |\tilde{U}_{w_1, e^{(\rho)}; w, w_4}|
 \chi[(w, w_1)\neq(e^{(\rho)},w_4)]
 \nonumber \\
 & \leq U
 \sum_{w_1 \in A}
 \sum_{w_4 \in (\Lambda\backslash A)\cup\{ w_1 \}}
 |\tilde{\psi}_{w_4, \uparrow}^{(w_1)}
 \tilde{\psi}_{w_4, \downarrow}^{(e^{(\rho)})}|
 \chi[w_4, w_1\neq e^{(\rho)}]
 +
 O(U \lambda^{-2}).
 \label{eq:tmp1}
\end{align}
Since $|\tilde{\psi}_{w_4, \uparrow}^{(w_1)}| \leq O(\lambda^{-1})$ for
$w_4 \neq w_1$, then (\ref{eq:tmp1}) is bounded by
\begin{equation}
 \leq U
 \sum_{w_1 \in A}
 |\tilde{\psi}_{w_1, \uparrow}^{(w_1)}
 \tilde{\psi}_{w_1, \downarrow}^{(e^{(\rho)})}|
 \chi[w_1\neq e^{(\rho)}] + O(U \lambda^{-1}).
 \label{eq:tmp2}
\end{equation}
Since $|\tilde{\psi}_{w_1, \uparrow}^{(e^{(\rho)})}| \leq O(\lambda^{-1})$ for
$w_1 \neq e^{(\rho)}$, we find the bound
\begin{align}
 & \sum_{w_1 \in A}
 \sum_{w \in \Lambda} \sum_{w_4 \in (\Lambda\backslash A)\cup\{ w_1 \}}
 |\tilde{U}_{w_1, e^{(\rho)}; w, w_4}|
 \chi[\overline{w=e^{(\rho)}, w_1=w_4}]
 \nonumber \\
 & \leq O(U \lambda^{-1}).
 \label{eq:D[general-A]}
\end{align}
From (\ref{eq:D[general-R]}) and (\ref{eq:D[general-A]}), we find the
bound (\ref{eq:bound_h_int[7]}).

\section{Proof of Theorem \ref{th:SW_UB} \label{sec:SW_UB}}

In this section, we show the proof of the upper bound of spin-wave
energy. The proof is given by the standard variational approach.
If we can show
\begin{align}
 \frac{(\Psi_{0,o}, H \Psi_{0,o})}{\|\Psi_{0,o}\|^2} \leq
  \frac{2U}{\lambda^4}
  \biggl[ &
   \frac{d(|q|+|q|^{-1})}{2}-
 \sum_{j=1}^{d} \cos \left( 2k\cdot e^{(j)} +\theta \right)+
   \frac{const}{\lambda^2}
  \biggr],
\end{align}
this concludes the Theorem \ref{th:SW_UB}. Here we evaluate 
this quantity.

Since $H_{\rm hop} \Psi_{0,o}=0$, we obtain
\begin{align}
 & \frac{(\Psi_{0,o}, H \Psi_{0,o})}{\|\Psi_{0,o}\|^2} =
 h_{\rm int}[\Psi_{0,o}, \Psi_{0,o}] +
 \nonumber \\
 & \sum_{\mu=0}^d \sum_{x \in \Lambda_o}
 \chi[(\mu, x) \neq (0,o)] h_{\rm int}[\Psi_{\mu, x}, \Psi_{0,o}]
 \frac{(\Psi_{0,o}, \Psi_{\mu, x})}{\|\Psi_{0,o}\|^2} +
 \nonumber \\
 & \sum_{\mu=0}^d \sum_{y \in\Lambda^\prime}
 \sum_{v, w \in \Lambda_o} \chi[v \neq w]
 h_{\rm int}[\Psi_{u, y, v, w}, \Psi_{0,o}]
 \frac{(\Psi_{0,o}, \Psi_{\mu, y, v, w})}{\|\Psi_{o,0}\|^2}.
\end{align}
Note $(\Psi_{0,o}, \Psi_{j, x})=0$ for $j=1, 2, \cdots, d$ because of the
representation of $a_{y, \sigma}$ with $y \in\Lambda^\prime$ in terms of
its dual
$$a_{y,\sigma}=\sum_{y^\prime \in \Lambda^\prime}C_{y, y^\prime;\sigma}b_{y^\prime,\sigma},$$
where $C_{y^\prime, y;\sigma}$ is a complex coefficient.
We also find $(\Psi_{0,o}, \Psi_{\mu, y, v, w})=0$ with the same
argument. Thus, we obtain
\begin{align}
 \frac{(\Psi_{0,o}, H \Psi_{0,o})}{\|\Psi_{0,o}\|^2}
 = h_{\rm int}[\Psi_{0,o}, \Psi_{0,o}] +
 \sum_{x \in \Lambda_o\setminus\{o\}}
 h_{\rm int}[\Psi_{0, x}, \Psi_{0,o}]
 \frac{(\Psi_{0,o}, \Psi_{0, x})}{\|\Psi_{0,o}\|^2}.
\end{align}
We already have the bounds of the matrix elements. We concentrate to
estimate the inner product between $\Psi_{0,o}$ and $\Psi_{0, x}$.
We represent $(\Psi_{0, o}, \Psi_{0, x})$
\begin{align}
 (\Psi_{0,o}, \Psi_{0,x})=\sum_{v, w \in \Lambda_o}
 e^{-i k \cdot (v-w)}
 (
 a_{v, \downarrow}^\dagger b_{v,\uparrow} \Psi_{\uparrow},
 a_{w, \downarrow}^\dagger b_{w+x, \uparrow} \Psi_{\uparrow}
 )
 \nonumber \\
 =
 \sum_{v, w \in \Lambda_o}
 e^{-i k \cdot (v-w)}
 \{b_{w+x,\uparrow}^\dagger, b_{v, \uparrow}\}
 \{a_{v, \downarrow}, a_{w,\downarrow}^\dagger\}
 \|\Psi_{\uparrow}\|^2
\end{align}
Using the relations
\begin{equation}
 \{a_{x, \sigma}, a_{y, \sigma}^\dagger\}
  = \sum_{w \in \Lambda} (\psi_{w, \sigma}^{(x)})^\ast
  \psi_{w,\sigma}^{(y)}
  \mbox{ and }
 \{b_{x, \sigma}^\dagger, b_{y, \sigma}\}
  = \sum_{w \in \Lambda} \tilde{\psi}_{w, \sigma}^{(x)}
  (\tilde{\psi}_{w,\sigma}^{(y)})^\ast
\end{equation}
and the explicit representation of $\psi_{x, \sigma}^{(y)}$, we obtain
\begin{align}
 & (\Psi_{0,o}, \Psi_{0,x})=L^d \sum_{w \in \Lambda}
 \biggl[
 \frac{\lambda^2+d(|q|^{1/2}+|q|^{-1/2})}{\lambda^2}
 \tilde{\psi}_{w, \uparrow}^{(x)}(\tilde{\psi}_{w,\uparrow}^{(o)})^\ast
 + 
 \nonumber \\
 & \frac{1}{\lambda^2}\sum_{j=1}^{d}
 \biggl(
 e^{-i k\cdot 2e^{(j)}+i \theta/2}
 \tilde{\psi}_{w, \uparrow}^{(x)}(\tilde{\psi}_{w,\uparrow}^{(2e^{(j)})})^\ast
 +
 e^{i k\cdot 2e^{(j)}-i \theta/2}
 \tilde{\psi}_{w, \uparrow}^{(x)}(\tilde{\psi}_{w,\uparrow}^{(-2e^{(j)})})^\ast
 \biggr)
 \biggr] \|\Phi_\uparrow\|^2.
\end{align}
To obtain this, we have also used the translational invariance of
$\tilde{\psi}_{x, \sigma}^{(y)}$: 
$\tilde{\psi}_{x, \sigma}^{(y)}=\tilde{\psi}_{x+w, \sigma}^{(y+w)}$ with
$w \in \Lambda_o$.
Thus we have
\begin{align}
 \left| \sum_{x \in \Lambda_o\backslash\{o\}} h[\Psi_{0,x}, \Psi_{0,o}]
 (\Psi_{0,o}, \Psi_{0,x}) \right|
 \leq L^d \frac{U}{\lambda^4} G_3 \sum_{x \in \Lambda_o\backslash\{o\}}
 |(\Psi_{0,o}, \Psi_{0,x})|
 \leq L^d \frac{U}{\lambda^6} K_1 \|\Psi_\uparrow\|^2,
\end{align}
where we use (\ref{eq:bound_h_int[3]}) and Lemma \ref{lemma:bound_sum_psi}.
The norm of $\Psi_{0, o}$ can be estimated as 
\begin{align}
 &
 \|\Psi_{0,o}\|^2 = L^d \sum_{w \in \Lambda}
 \biggl[
 \frac{\lambda^2+d(|q|^{1/2}+|q|^{-1/2})}{\lambda^2}
 \tilde{\psi}_{w, \uparrow}^{(o)}(\tilde{\psi}_{w,\uparrow}^{(o)})^\ast
 +
 \nonumber \\
 &
 \frac{1}{\lambda^2} \sum_{j=1}^d
 \biggl(
 e^{-i k\cdot 2e^{(j)}+i\theta/2}
 \tilde{\psi}_{w, \uparrow}^{(o)}(\tilde{\psi}_{w,\uparrow}^{(2e^{(j)})})^\ast
 -
 e^{i k\cdot 2e^{(j)}-i\theta/2}
 \tilde{\psi}_{w, \uparrow}^{(o)}(\tilde{\psi}_{w,\uparrow}^{(-2e^{(j)})})^\ast
 \biggr)
 \biggr] \|\Phi_{\uparrow}\|^2
 \geq L^d K_2 \|\Psi_\uparrow\|.
\end{align}
Then we have
\begin{align}
 \frac{(\Psi_{0,o}, H \Psi_{0,o})}{\|\Psi_{0,o}\|^2}
 \leq & |h[\Psi_{0,o}, \Psi_{0,o}]| + \frac{U}{\lambda^6} \frac{K_1}{K_2}
 \nonumber \\
 \leq &
 \frac{2U}{\lambda^4}
 \biggl[
 \frac{d(|q|+|q|^{-1})}{2} -
 \sum_{j=1}^d \cos (k \cdot 2e^{(j)}+\theta)
 +\frac{K_1/K_2}{\lambda^2}
 \biggr].
\end{align}
This completes the proof. \qed

\paragraph{Acknowledgments}

I would like to thank T. Koma, M. Oshikawa, A. Tanaka and H. Tasaki for
helpful suggestions and great encouragements. I also thank to C. Itoi
for carefully reading the manuscript and kind suggestions.

\appendix

\setcounter{equation}{0}
\renewcommand{\theequation}{\Alph{section}.\arabic{equation}}

\section{Localized Basis \label{A:localized_bases}}

\subsection{Single Electron Problem}

First, we consider a single-electron problem with periodic boundary
condition to construct the localized basis. A single electron state with
spin $\sigma$: $\Phi_{\sigma}$ can be written in a form
\begin{equation}
 \Phi_{\sigma} =
  \sum_{x \in \Lambda}
  \psi_{x, \sigma} c_{x, \sigma}^\dagger \Psi_{\rm vac}.
\end{equation}
Since $H_{\rm int} \Phi_{\sigma} = 0$ is obvious, the
Schr\"{o}dinger equation
$H \Phi_{\sigma} = \varepsilon^{(\sigma)} \Phi_{\sigma}$
is reduced to
\begin{equation}
 \sum_{x \in \Lambda} t_{x, y}^{(\sigma)} \psi_{y, \sigma}
  = \varepsilon^{(\sigma)} \psi_{x, \sigma}.
\end{equation}
From the representation of $H_{\rm hop}$:
\begin{align}
 H_{\rm hop} =
 t \sum_{x \in \Lambda_o}
 \sum_{\sigma = \uparrow, \downarrow}
 \biggl[
 d (|q|^{1/2} + |q|^{-1/2}) n_{x, \sigma}
 +
 \nonumber \\
 \sum_{j = 1}^{d}
 \left(
 e^{i p(\sigma) \frac{\theta}{2}}
 c_{x, \sigma}^\dagger c_{x + 2 e^{(j)}, \sigma}
 + e^{- i p(\sigma) \frac{\theta}{2}}
 c_{x + 2 e^{(j)}, \sigma}^\dagger c_{x, \sigma}
 \right)
 \biggr]
 \nonumber \\
 +
 \lambda t \sum_{j = 1}^{d} \sum_{x \in \Lambda_j}
 \sum_{\sigma = \uparrow, \downarrow}
 \biggl[
 \lambda n_{x, \sigma}
 + q^{- p(\sigma)/4}
 c_{x, \sigma}^\dagger c_{x - e^{(j)}, \sigma}
 +
 \left( q^{- p(\sigma)/4} \right)^{\ast}
 c_{x - e^{(j)}, \sigma}^\dagger c_{x, \sigma}
 \nonumber \\
 + q^{p(\sigma)/4}
 c_{x, \sigma}^\dagger c_{x + e^{(j)}, \sigma}
 + \left( q^{p(\sigma)/4} \right)^{\ast}
 c_{x + e^{(j)}, \sigma}^\dagger c_{x, \sigma}
 \biggr],
\end{align}
we obtain an expression of the equations
\begin{equation}
 \varepsilon \psi_{x, \sigma}
  =
  \begin{cases}
   \displaystyle
   t d (|q|^{1/2} + |q|^{-1/2})
   \psi_{x, \sigma}
   + \\
   \displaystyle
   \quad t \sum_{j = 1}^{d}
   \biggl[
   e^{i p(\sigma) \frac{\theta}{2}} \psi_{x + 2 e^{(j)}, \sigma}
   + e^{- i p(\sigma) \frac{\theta}{2}} \psi_{x - 2 e^{(j)}, \sigma}
   +
   \\
   \hfill
   \lambda \left( q^{- p(\sigma)/4} \right)^{\ast}
   \psi_{x + e^{(j)}, \sigma}
   + \lambda \left( q^{p(\sigma)/4} \right)^{\ast}
   \psi_{x - e^{(j)}, \sigma}
   \biggr]
   & \mbox{if } x \in \Lambda_o
   \\
   \displaystyle
   t \lambda^2 \psi_{x, \sigma}
   + t \lambda
   \left[
   q^{- p(\sigma)/4} \psi_{x - e^{(j)}, \sigma}
   + q^{p(\sigma)/4} \psi_{x + e^{(j)}, \sigma}
   \right]
   & \mbox{if } x \in \Lambda^\prime
  \end{cases}
  . \label{eq:schroedinger_eq}
\end{equation}
Since $t_{x+w, y+w}^{(\sigma)} = t_{x, y}^{(\sigma)}$ for
$w \in \Lambda_o$, we can use the Bloch's theorem which ensures that
$\psi_{x, \sigma}$ can be written in a form
\begin{equation}
 \psi_{x, \sigma} = e^{i k x} v_{x, \sigma}(k)
  \label{eq:Bloch_decomp}
\end{equation}
where $v_{x, \sigma}(k)$ satisfies
\begin{equation}
 v_{x+w, \sigma}(k) = v_{x, \sigma}(k)
\end{equation}
for $w \in {\mathbb Z}$, and $k \in \K$ is a wave number vector
(see eq. (\ref{eq:def_of_K})).
We define a mapping $u$ by
\begin{equation}
 u(x) =
  \begin{cases}
   0 & \mbox{if} \quad x \in \Lambda_o \\
   j & \mbox{if} \quad x \in \Lambda_j
  \end{cases}
  .
\end{equation}
We rewrite eq. (\ref{eq:Bloch_decomp}) to a form
\begin{equation}
 \psi_{x, \sigma} = e^{i k \cdot x} v_{u(x), \sigma}(k),
  \label{eq:Bloch_decomp2}
\end{equation}
since $v_{x, \sigma}(k)$ is translationally invariant.
%
%
By substituting the representation (\ref{eq:Bloch_decomp2}) into the
Schr\"{o}dinger equation (\ref{eq:schroedinger_eq}), we find the
equation
\begin{equation}
 \varepsilon(k) v_\sigma(k) = {\sf M}^{(\sigma)}(k)  v_\sigma(k)
\end{equation}
where $v_\sigma(k)$ is a $(d+1)$-dimensional complex vector defined by
$v_\sigma(k)  = (v_{\mu, \sigma}(k))_{\mu = 0}^{d}$ and
${\sf M}^{(\sigma)}(k) =(M_{\mu, \rho}^{(\sigma)}(k))_{\mu, \rho =0}^d$ is
a $(d+1)\times(d+1)$ matrix. $M_{\mu, \rho}^{(\sigma)}(k)$ is defined by
\begin{equation}
 M_{\mu, \rho}^{(\sigma)}(k) :=
  \sum_{y \in \Lambda_\rho} t_{x, y}^{(\sigma)}e^{i k\cdot(y-x)}
\end{equation}
for
$u(x)=\mu$ and $u(y)=\rho$. Here we define $\Lambda_\rho = \Lambda_o$
for $\rho=0$ and $\Lambda_\rho = \Lambda_j$ for $\rho=j$ for
convenience. We use indices $\mu$ and $\rho$ as an integer of
$0, 1, 2, \cdots, d$, and $j$ and $l$ as an integer of $1, 2, \cdots, d$.
\begin{equation}
 \frac{M_{\mu, \rho}^{(\sigma)}(k)}{t}
  =
  \begin{cases}
   \displaystyle
   d (|q|^{1/2} + |q|^{-1/2})
   + 2 \sum_{j = 1}^{d}
   \cos \left[ \frac{\theta}{2} p(\sigma) + 2 k \cdot e^{(j)} \right]
   &\mu = \rho = 0 \\
   \displaystyle
   \lambda
   \left[
   \left( q^{- p(\sigma)/4} \right)^{\ast} e^{i k \cdot e^{(j)}}
   + \left( q^{p(\sigma)/4} \right)^{\ast} e^{- i k \cdot e^{(j)}}
   \right]
   & \mu = 0, \rho = j \\
   \lambda^2
   & \mu = \rho = j \\
   0
   & \mbox{otherwise}
  \end{cases}
  .
\end{equation}
We denote the energy eigenvalues as
$\varepsilon_\mu^{(\sigma)}(k)$, 
where we assigned index $\mu$ such that
$\varepsilon_\mu^{(\sigma)}(k) \leq \varepsilon_{\mu+1}^{(\sigma)}(k)$.
We can easily calculate the eigenvalues of the matrix
${\sf M}^{(\sigma)}(k)$:
\begin{equation}
 \frac{\varepsilon_\mu^{(\sigma)}(k)}{t} =
 \begin{cases}
  0 
  & \mbox{ for } \mu=0
  \\
  \lambda^2
  & \mbox{ for } \mu=1, 2, \cdots, d-1
  \\
  \displaystyle
  \lambda^2 + d (|q|^{1/2} + |q|^{-1/2})
  + 2 \sum_{j = 1}^{d}
  \cos \left[ \frac{\theta}{2} p(\sigma) + k_j \right]
  & \mbox{ for } \mu=d
 \end{cases}
 .
\end{equation}

We define $v_\sigma^{(0)}(k)=(v_{\mu, \sigma}^{(0)}(k))_{\mu=0}^d$ by
\begin{equation}
 v_{\mu, \sigma}^{(0)}(k) =
  \begin{cases}
   1  
   & \mbox{if } \mu = 0\\
   \displaystyle
   -\frac{1}{\lambda}
   (
   q^{-p(\sigma)/4} e^{-i k \cdot e^{(j)}}
   + q^{p(\sigma)/4} e^{i k \cdot e^{(j)}}
   )
   & \mu = j (=1, 2, \cdots, d)
  \end{cases}.
\end{equation}
This vector is an eigenvector which belongs to the zero-energy eigenvalue
of ${\sf M}^{(\sigma)}(k)$.
We also define $v_\sigma^{(j)}(k)=(v_{\mu, \sigma}^{(j)}(k))_{\mu=0}^d$ for
each $j$ by
\begin{equation}
 v_{\mu, \sigma}^{(j)}(k) =
  \begin{cases}
   \displaystyle
   \frac{1}{\lambda}
   \{
   (q^{-p(\sigma)/4})^{\ast} e^{i k \cdot e^{(j)}}
   + (q^{p(\sigma)/4})^{\ast} e^{-i k \cdot e^{(j)}}
   \}
   & \mbox{if} \quad \mu = 0
   \\
   1 
   & \mbox{if} \quad \mu = j
   \\
   0 
   & \mbox{otherwise}
  \end{cases}
  .
\end{equation}
This is orthogonal to $v_\sigma^{(0)}(k)$. The vectors
$\{ v_\sigma^{(\mu)}(k) \}_{\mu=0}^d$ for fixed $k$ and $\sigma$ is a
basis of ${\mathbb C}^{d+1}$, since $v_\sigma^{(\mu)}(k)$ with
$\mu = 0, 1, \cdots, d$ are linearly independent of each other.

We also introduce the dual of the basis
$\{ v_\sigma^{(\mu)}(k) \}_{\mu=0}^d$. We define the Gramm matrix
${\sf G}^{(\sigma)}(k)=( G_{\mu, \rho}^{(\sigma)}(k))_{\mu, \rho=0}^d$
by \cite{fn:inner-product}
\begin{equation}
 G_{\mu, \rho}^{(\sigma)}(k)
  = ( v_\sigma^{(\mu)}(k), v_\sigma^{(\rho)}(k)).
\end{equation}
Since the vectors $v_\sigma^{(\mu)}(k)$ with $\mu=0, 1, \cdots, d$ are
linearly independent of each other, the corresponding Gramm matrix is
invertible. We define the dual vectors $\tilde{v}_\sigma^{(\mu)}(k)$ by
\begin{equation}
 \tilde{v}_\sigma^{(\mu)}(k) =
  \sum_{\rho = 0}^{d} v_\sigma^{(\rho)}(k)
  {G_{\rho, \mu}^{(\sigma)}(k)}^{-1}
\end{equation}
for each $\mu$. They satisfy
\begin{equation}
 ( \tilde{v}_\sigma^{(\mu)}(k), v_\sigma^{(\rho)}(k) )
  = \delta_{\mu, \rho}
  \mbox{ and }
  \sum_{\eta = 0}^{d}
  (\tilde{v}_{\mu, \sigma}^{(\eta)}(k))^{\ast} v_{\rho, \sigma}^{(\eta)}(k)
  = \delta_{\mu, \rho}.
\end{equation}

\subsection{Construction of the Localized Basis}

We define $\psi_{y, \sigma}^{(x)}$ and
$\tilde{\psi}_{y, \sigma}^{(x)}$ by
\begin{equation}
 \psi_{y, \sigma}^{(x)} :=
  \frac{1}{L^d} \sum_{k \in \K} e^{i k \cdot (y -x)}
  v_{u(y), \sigma}^{(u(x))}(k)
\end{equation}
and
\begin{equation}
  \tilde{\psi}_{y, \sigma}^{(x)} =
  \frac{1}{L^{d}} \sum_{k \in \K}
  e^{i k \cdot (y - x)} \tilde{v}_{u(y), \sigma}^{(u(x))}(k)
  .
\end{equation}
They satisfy
\begin{align}
 \sum_{w \in \Lambda}
 (\tilde{\psi}_{w, \sigma}^{(x)})^{\ast} \psi_{w, \sigma}^{(y)}
 = \delta_{x, y}
 \quad \mbox{and} \quad
 \sum_{w \in \Lambda}
 (\tilde{\psi}_{x, \sigma}^{(w)})^{\ast} \psi_{y, \sigma}^{(w)}
 = \delta_{x, y}.
\end{align}
We can easily obtain explicit representation of
$\psi_{y, \sigma}^{(x)}$:
\begin{equation}
 \psi_{y, \sigma}^{(x)} =
   \begin{cases}
    \displaystyle
    -\frac{q^{p(\sigma)/4}}{\lambda}
    \sum_{j=1}^{d} \delta_{x - e^{(j)}, y}
    + \delta_{x, y}
    -\frac{q^{-p(\sigma)/4}}{\lambda}
    \sum_{j=1}^{d} \delta_{x + e^{(j)}, y}
    & \mbox{ if } x \in \Lambda_o \\
    \displaystyle
    \frac{(q^{-p(\sigma)/4})^\ast}{\lambda}
    \delta_{x - e^{(j)}, y}
    + \delta_{x, y}
    + \frac{(q^{p(\sigma)/4})^\ast}{\lambda}
    \delta_{x + e^{(j)}, y}
    & \mbox{ if } x \in \Lambda_j
   \end{cases}
   .
\end{equation}

\subsection{Calculation of $\tilde{v}_{\sigma}^{(\mu)}(k)$}

To estimate $\tilde{\psi}_{y, \sigma}^{(x)}$, we obtain the explicit
representation of $\tilde{v}_{\sigma}^{(\mu)}(k)$. We write
${\sf G}^{(\sigma)}(k)$ in the form
\begin{equation}
 {\sf G}^{(\sigma)}(k) =
  \begin{pmatrix}
   |v_\sigma^{(0)}(k)|^2 & 0 \hfill  \cdots \hfill 0 \\
   0 & \\
   \vdots & \displaystyle \unit + \frac{1}{\lambda^2} {\bm w}_\sigma(k)  \otimes {\bm w}_\sigma(k)^\ast \\
   0 &
  \end{pmatrix}
\end{equation}
where $\unit$ is the $d \times d$ identity matrix, and
${\bm w}_\sigma(k)$ and ${\bm w}_\sigma(k)^{\ast}$ are $d$-dimensional
vectors defined by
\begin{equation}
 {\bm w}_\sigma(k) =
  \left(
   (q^{-p(\sigma)/4})^{\ast} e^{i k \cdot e^{(j)}}
   + (q^{p(\sigma)/4})^{\ast} e^{-i k \cdot e^{(j)}}
  \right)_{j = 1}^{d}
\end{equation}
and
\begin{equation}
 {\bm w}_\sigma(k)^{\ast} =
  \left(
   q^{-p(\sigma)/4} e^{-i k \cdot e^{(j)}}
   + q^{p(\sigma)/4} e^{i k \cdot e^{(j)}}
  \right)_{j = 1}^{d}.
\end{equation}
$\otimes$ denotes the standard tensor product.
There is a general formula to obtain the inverse matrix
\begin{equation}
 \left(
  \unit + \frac{1}{\lambda^2}
  {\bm w}_\sigma(k) \otimes {\bm w}_\sigma(k)^{\ast}
 \right)^{-1}
  = \unit
  - \frac{1}{|{\bm w}_\sigma(k)|^2 + \lambda^2}
  {\bm w}_\sigma(k) \otimes {\bm w}_\sigma(k)^{\ast}.
\end{equation}
Then, we obtain the explicit form of ${{\sf G}^{(\sigma)}(k)}^{-1}$
\begin{equation}
 {{\sf G}^{(\sigma)}(k)}^{-1} =
  \begin{pmatrix}
   |v_\sigma^{(0)}(k)|^{-2} & 0 \hfill \cdots \hfill 0 \\
   0 &  \\
   \vdots & \displaystyle
   \unit -
   \frac{1}{|{\bm w}_\sigma(k)|^2 + \lambda^2}
   {\bm w}_\sigma(k) \otimes {\bm w}_\sigma(k)^{\ast} \\
   0 &
  \end{pmatrix}
\end{equation}
Thus, we find an explicit representation of
$\tilde{v}_\sigma^{(\mu)}(k)$:
\begin{align}
 & \tilde{v}_{\mu, \sigma}^{(0)}(k)=
  \frac{1}{1 + \frac{S_1}{\lambda^2} + \frac{S_2^{(\sigma)}(k)}{\lambda^2}}
 \begin{cases}
   \displaystyle
   1
   & \mbox{ for } \mu = 0 \\
   \displaystyle
   - \frac{1}{\lambda}
   \left(
   q^{-p(\sigma)/4} e^{-i k \cdot e^{(j)}}
   + q^{p(\sigma)/4} e^{i k \cdot e^{(j)}}
   \right)
   & \mbox{ for } \mu = j
 \end{cases},
\end{align}
\begin{align}
 & \tilde{v}_{\mu, \sigma}^{(j)}(k)
 =v_{\mu, \sigma}^{(j)}(k) +
 \frac{(q^{-p(\sigma)/4})^{\ast} e^{i k \cdot e^{(j)}}
 + (q^{p(\sigma)/4})^{\ast} e^{-i k \cdot e^{(j)}}}
 {\lambda^2 + S_1 + S_2^{(\sigma)}(k)} \times
 \nonumber \\
 &
  \begin{cases}
   \displaystyle
   \frac{S_1 + S_2^{(\sigma)}(k)}{\lambda}
   & \mbox{for} \quad \mu=0 \\
   \displaystyle
   q^{-p(\sigma)/4} e^{- i k \cdot e^{(l)}}
   + q^{p(\sigma)/4} e^{i k \cdot e^{(l)}}
   & \mbox{for} \quad \mu=l(=1, 2, \cdots, d)
  \end{cases}
 ,
\end{align}
where
\begin{equation}
 S_1 = d (|q|^{1/2} + |q|^{-1/2}),
  \quad
  S_2^{(\sigma)}(k) = 2 \sum_{j = 1}^{d} \cos
  \left(
   {2 k \cdot e^{(j)} + \frac{\theta}{2}p(\sigma)}
  \right).
\end{equation}

\subsection{Proof of Lemma \ref{lemma:bound_sum_psi} \label{A:estimates_of_psi}}

Here we prove Lemma \ref{lemma:bound_sum_psi}.
We give the estimates of $\tilde{\psi}_{x, \sigma}^{(y)}$ only for
$x, y \in \Lambda_o$, since the estimates for other $x, y \in \Lambda$
are almost the same as that for $x, y \in \Lambda_o$.

From the definition of $\tilde{\psi}_{x, \sigma}^{(y)}$, we can represent
\begin{align}
 \tilde{\psi}_{x, \sigma}^{(y)}
  = &
  \frac{1}{L^{d}} \sum_{k \in \K}
  e^{i k \cdot (x - y)} \tilde{v}_{0, \sigma}^{(0)}(k)
  = \frac{1}{L^{d}} \sum_{k \in \K}
  e^{i k \cdot (x - y)}
  \frac{1}{1 + \frac{S_1}{\lambda^2} + \frac{S_2^{(\sigma)}(k)}{\lambda^2}}
 \nonumber \\
  = &
 \frac{1}{1+\frac{S_1}{\lambda^2}}
  \frac{1}{L^{d}} \sum_{k \in \K}
  e^{i k \cdot (x - y)}
  \frac{1}{1 + \frac{S_2^{(\sigma)}(k)}{\lambda^2 + S_1}}
\end{align}
We can expand the last factor, since $|S_2^{(\sigma)}(k)| \leq 2d < \lambda^2+S_1$.
Then, we obtain
\begin{align}
 \tilde{\psi}_{x, \sigma}^{(y)}
  = &
  \frac{1}{1 + \frac{S_1}{\lambda^2}}
  \sum_{n = 0}^{\infty}
  \frac{1}{L^{d}} \sum_{k \in \K} e^{i k \cdot (x - y)}
  \left( -\frac{S_2^{(\sigma)}(k)}{\lambda^2 + S_1} \right)^{n}
 \nonumber \\
  = &
  \frac{1}{1 + \frac{S_1}{\lambda^2}}
  \sum_{n = 0}^{\infty} \left( -\frac{1}{\lambda^2 + S_1} \right)^{n}
  \frac{1}{L^{d}} \sum_{k \in \K} e^{i k \cdot (x - y)}
  {S_2^{(\sigma)}(k)}^n
\end{align}
Here,
\begin{align}
 & \frac{1}{L^{d}} \sum_{k \in \K} e^{i k \cdot (x - y)}
 {S_2^{(\sigma)}(k)}^{n}
 =
 \frac{1}{L^{d}} \sum_{k \in \K} e^{i k \cdot (x - y)}
 \left[
 \sum_{j = 1}^{d}
 \left(
 e^{2 i k \cdot e^{j} + \frac{\theta p(\sigma)}{2}}
 + e^{-2 i k \cdot e^{j} - \frac{\theta p(\sigma)}{2}}
 \right)
 \right]^{n}
 \nonumber \\
 & =
 \sum_{j_1, j_2, \cdots, j_n = 1}^{d}
 \sum_{s_1, s_2, \cdots, s_n = \pm 1}
 \frac{1}{L^{d}} \sum_{k \in \K}
 e^{i k \cdot (x - y + 2 \sum_{l = 1}^{n} s_l k \cdot e^{(j_{l})})
 + i \sum_{l = 1}^{n}s_l \frac{\theta p(\sigma)}{2}}
 \nonumber \\
 & =
 \sum_{j_1, j_2, \cdots, j_n = 1}^{d}
 \sum_{s_1, s_2, \cdots, s_n = \pm 1}
 e^{i \sum_{l = 1}^{n}s_l \frac{\theta p(\sigma)}{2}}
 \delta_{x + 2 \sum_{l = 1}^{n} s_l k \cdot e^{(j_{l})}, y}
 .
 \label{eq:S_2^n}
\end{align}
Thus, we find a bound
\begin{align}
 \sum_{x \in \Lambda_o} |\tilde{\psi}_{x, \sigma}^{(y)}|
 \leq &
 \frac{1}{1 + \frac{S_1}{\lambda^2}}
 \sum_{n = 0}^{\infty}
 \left( \frac{1}{\lambda^2 + A} \right)^n
 \sum_{x \in \Lambda_o}
 \left|
 \frac{1}{L^{d}}
 \sum_{k \in \K} e^{i k \cdot (x - y)} {S_2^{(\sigma)}(k)}^{n}
 \right|
 \nonumber \\
 = &
 \frac{1}{1 + \frac{S_1}{\lambda^2}}
 \sum_{n = 0}^{\infty}
 \left( \frac{1}{\lambda^2 + A} \right)^n
 \sum_{x \in \Lambda_o}
 \sum_{j_1, j_2, \cdots, j_n = 1}^{d}
 \sum_{s_1, s_2, \cdots, s_n = \pm 1}
 \delta_{x + 2 \sum_{l = 1}^{n} s_l k \cdot e^{(j_{l})}, y}
 \nonumber \\
 = &
 \frac{1}{1 + \frac{S_1}{\lambda^2}}
 \sum_{n = 0}^{\infty}
 \left( \frac{2d}{\lambda^2 + A} \right)^n
 =
 \frac{1}{1 + \frac{S_1}{\lambda^2} - \frac{2d}{\lambda^2}}.
\end{align}
We can represent
$\tilde{\psi}_{x, \sigma}^{(y)} - \psi_{x, \sigma}^{(y)}$ in terms of
$\tilde{\psi}_{w, \sigma}^{(y)}$ where $w \in \Lambda_o$ with
$|w-x|\leq 1$,
\begin{align}
 \tilde{\psi}_{x, \sigma}^{(y)} - \psi_{x, \sigma}^{(y)}
 = &
 -
 \frac{1}{L^{d}} \sum_{k \in \K}
 e^{i k \cdot (x - y)}
 \frac{S_1 + S_2^{(\sigma)}(k)}{\lambda^2}
 \frac{1}{1 + \frac{S_1}{\lambda^2} + \frac{S_2^{(\sigma)}(k)}{\lambda^2}}
 \nonumber \\
 =
 -\frac{S_1}{\lambda^2} & \tilde{\psi}_{x, \sigma}^{(y)}
 -\frac{1}{\lambda^2} \sum_{j=1}^d \sum_{s=\pm 1}
 e^{i s\frac{\theta p(\sigma)}{2}}
 \tilde{\psi}_{x+2s e^{(j)}, \sigma}^{(y)}.
 \label{eq:tilde_psi-psi}
\end{align}
Thus we find the following bound
\begin{equation}
 \sum_{x \in \Lambda_o}|\tilde{\psi}_{x, \sigma}^{(y)}-\psi_{x, \sigma}^{(y)}|
  \leq \frac{\mbox{const}}{\lambda^2}
\end{equation}
For other $x, y \in \Lambda$, we can write
$\tilde{\psi}_{y, \sigma}^{(x)}$ in terms of
$\tilde{\psi}_{v, \sigma}^{(w)}$ with $v, w \in \Lambda_o$ like the
representation (\ref{eq:tilde_psi-psi}) and can obtain the bounds.
Then we can conclude the lemma \ref{lemma:bound_sum_psi}. \qed

\subsection{Proof of Lemma \ref{lemma:bound_psi}}

Here, we give proof of Lemma \ref{lemma:bound_psi}. We only show the
proof for the case of $x, y \in \Lambda_o$, since the estimates for other
$x, y \in \Lambda$ are almost the same as that for $x, y \in \Lambda_o$.

In the eq. (\ref{eq:S_2^n}), if $\|y - x \|_1 > n$, then
$\delta_{y, x + 2\sum_{l = 1}^{n} s_l e^{(j_l)}}=0$. Thus we find
\begin{align}
 |\tilde{\psi}_{x}^{(y)}|
 \leq &
 \frac{1}{1 + \frac{S_1}{\lambda^2}}
 \sum_{n = \| x- y \|_1}^{\infty}
 \left( \frac{1}{\lambda^2 + S_1} \right)^{n}
 \sum_{j_1, j_2, \cdots, j_n = 1}^{d}
 \sum_{s_1, s_2, \cdots, s_n = \pm 1}
 \delta_{x + 2 \sum_{l = 1}^{n} s_l k \cdot e^{(j_{l})}, y}
 \nonumber \\
 \leq &
 \frac{1}{1 + \frac{S_1}{\lambda^2}}
 \sum_{n = \| x- y \|_1}^{\infty}
 \left( \frac{2d}{\lambda^2 + S_1} \right)^{n}
 \nonumber \\
 = &
 \left( \frac{2d}{\lambda^2 + S_1} \right)^{\|x - y\|_1}
 \frac{1}{1 + \frac{S_1}{\lambda^2} - \frac{2d}{\lambda^2}}
\end{align}
For other $x, y \in \Lambda$, we can find similar bounds. Then, we obtain
the first inequality in Lemma \ref{lemma:bound_psi}.

We also obtain (\ref{eq:bound_diff_psi00}) and
(\ref{eq:bound_diff_psi01}) from the representations
\begin{align}
 \tilde{\psi}_o^{(o)} - \psi_o^{(o)}
 =
 - \frac{1}{\lambda^2} \frac{1}{L^d} \sum_{k \in \K}
 \frac{S_1 + S_2^{(\sigma)}(k)}{1 + \frac{S_1}{\lambda^2} + \frac{S_2^{(\sigma)}(k)}{\lambda^2}}
 =
 -\frac{S_1}{\lambda^2} \tilde{\psi}_o^{(o)}
 -\frac{1}{\lambda^2} \sum_{j=1}^{d} \sum_{s = \pm 1}
 e^{i s \frac{\theta p(\sigma)}{2}} \tilde{\psi}_{2 e^{(j)}}^{(o)}
\end{align}
and
\begin{align}
 \tilde{\psi}_{\pm e^{(j)}}^{(o)} - \psi_{\pm e^{(j)}}^{(o)}
 =
 -\frac{S_1}{\lambda^2} \tilde{\psi}_{\pm e^{(j)}}^{(o)}
 - \frac{1}{\lambda^2} \sum_{l=1}^{d}
 \sum_{s = \pm 1}
 e^{i s \frac{\theta p(\sigma)}{2}}
 \tilde{\psi}_{x + 2e^{(l)} \pm e^{(j)}}^{(o)}.
\end{align}
They conclude the Lemma \ref{lemma:bound_psi}. \qed

\setcounter{equation}{0}

\section{Proof of Lemma \ref{lemma:SW-1} \label{A:proof_SW-1}}

Let $\Phi_{0} \in {\cal H}_{k}$ be the eigenstate which belongs to the
lowest energy eigenvalue $\tilde{E}_0$. We can expand $\Phi_{0}$ by
the basis $\B_k$:
\begin{equation}
 \Phi_{0} = \sum_{\Psi \in {\cal B}_{k}} C(\Psi) \Psi.
\end{equation}
The eigenvalue equation is 
\begin{equation}
  \sum_{\Psi, \Phi \in {\cal B}_{k}}
  C(\Psi) h[\Phi, \Psi]\Phi
  =
  \sum_{\Phi \in {\cal B}_{k}}
  \tilde{E}_0 C(\Phi) \Phi.
\end{equation}
Thus, we have an equation for the coefficient 
\begin{equation}
 \tilde{E}_{0} C(\Phi)
  =
  \sum_{\Psi \in {\cal B}_{k}}
  h[\Phi, \Psi] C(\Psi)
\end{equation}
for each $\Phi \in \B_k$. Then, we obtain
\begin{equation}
 \tilde{E}_0
  =
  h[\Phi, \Phi]
  +
  \sum_{\Psi \in {\cal B}_{k}\backslash \{ \Phi \}}
  h[\Phi, \Psi] \frac{C(\Psi)}{C(\Phi)}
\end{equation}
Since $\tilde{E}_0
 \in \mathbb{R}$, the real part of the equation is
\begin{align}
 \tilde{E}_0
 =&
 \Re
 \left[
 h[\Phi, \Phi]
 +
 \sum_{\Psi \in {\cal B}_{k}\backslash \{ \Phi \}}
 h[\Phi, \Psi] \frac{C(\Psi)}{C(\Phi)}
 \right]
 \nonumber \\
 =&
 \Re[h[\Phi, \Phi]]
 +
 \sum_{\Psi \in {\cal B}_{k}\backslash \{ \Phi \}}
 \cos \left(\theta_{\Phi, \Psi} \right)
 |h[\Phi, \Psi]|
 \left|
 \frac{C(\Psi)}{C(\Phi)}
 \right|
 \nonumber \\
 \geq&
 \Re[h[\Phi, \Phi]]
 -
 \sum_{\Psi \in {\cal B}_{k}\backslash \{ \Phi \}}
 |h[\Phi, \Psi]|
 \left|
 \frac{C(\Psi)}{C(\Phi)}
 \right|,
\end{align}
where $\theta_{\Phi, \Psi} \in \mathbb{R}$ gives the phase factor of
$h[\Phi, \Psi] C(\Psi)/C(\Phi)$. If we choose $\Phi$ such that
$|C(\Phi)| = \max_{\Phi \in \B_k} |C(\Psi)|$, {\it i.e.}
$|C(\Psi)/C(\Phi)| \leq 1$ for $\forall \Psi \in {\cal B}_{k}$,
then we find
\begin{equation}
 \tilde{E}_0
  \geq
  \Re[h[\Phi, \Phi]]
  -
  \sum_{\Psi \in {\cal B}_{k}\backslash \{ \Phi \}}
  |h[\Phi, \Psi]|
  =
  D[\Phi].
\end{equation}
Since $D[\Phi] \geq \min_{\Psi \in {\cal B}_{k}} D[\Psi]$, we have
proved lemma \ref{lemma:SW-1}. \qed


\begin{thebibliography}{10}
\bibitem{fn:inner-product}
	 We denote $(v, v^\prime)$ inner product of $(d+1)$-dimensional
	 vectors $v$ and $v^\prime$, which is defined by
	 $(v, v^\prime)=\sum_{\mu=0}^d (v_\mu)^\ast v_\mu^\prime$.
\end{thebibliography}
\end{document}